\documentclass[useAMS,usenatbib]{mn2e}
\usepackage{epsfig}
\usepackage{amssymb}
\usepackage{amsmath}
\usepackage{lscape}
\usepackage{longtable}
\usepackage[normal]{caption}
\usepackage{appendix}
\usepackage{bm}

\title[AGN and star formation across cosmic time]{AGN and star formation across cosmic time}

\author[M.~Symeonidis and M. J. ~Page] 
{\parbox{\textwidth}{\raggedright
M.~Symeonidis,$^{1}$\thanks{E-mail: \texttt{m.symeonidis@ucl.ac.uk}}
and M. J. ~Page,$^{1}$ }\vspace{0.4cm}\\
\parbox{\textwidth}{\raggedright $^{1}$ Mullard Space Science
  Laboratory, University College London, Holmbury St. Mary, Dorking,
  Surrey RH5 6NT, UK}}

\begin{document}

\date{Accepted  Received; in original form}

\pagerange{\pageref{firstpage}--\pageref{lastpage}} \pubyear{2014}

\maketitle

\label{firstpage}

\begin{abstract}
We investigate the balance of power between stars and AGN across cosmic history, based on the comparison between the infrared (IR) galaxy luminosity function (LF) and the IR AGN LF. The former corresponds to emission from dust heated by stars and AGN, whereas the latter includes emission from AGN-heated dust only. 
We find that at all redshifts (at least up to z$\sim$2.5), the high luminosity tails of the two LFs converge, indicating that the most infrared-luminous galaxies are AGN-powered. Our results shed light to the decades-old conundrum regarding the flatter high-luminosity slope seen in the IR galaxy LF compared to that in the UV and optical. 
We attribute this difference to the increasing fraction of AGN-dominated galaxies with increasing total infrared luminosity ($L_{\rm IR}$).
We partition the $L_{\rm IR} - z$ parameter space into a star-formation and an AGN-dominated region, finding that the most luminous galaxies at all epochs lie in the AGN-dominated region. This sets a potential `limit' to attainable star formation rates, casting doubt on the abundance of `extreme starbursts': if AGN did not exist, $L_{\rm IR}>10^{13}$\,L$_{\odot}$ galaxies would be significantly rarer than they currently are in our observable Universe. We also find that AGN affect the average dust temperatures ($T_{\rm dust}$) of galaxies and hence the shape of the well-known $L_{\rm IR}-T_{\rm dust}$ relation. We propose that the reason why local ULIRGs are hotter than their high redshift counterparts is because of a higher fraction of AGN-dominated galaxies amongst the former group. 
\end{abstract}


\section{Introduction}
\label{sec:introduction}

In star-forming galaxies a significant fraction of the stellar UV and optical radiation is absorbed by dust and re-emitted in the infrared (IR). As a result infrared emission is commonly used as a proxy for star-formation and there exists a set of straight-forward, widely used calibrations for converting total IR luminosity ($L_{\rm IR}$, 8--1000$\mu$m) to the star-formation rate (SFR; e.g. Kennicutt 1998\nocite{Kennicutt98}; et al. 2009\nocite{Kennicutt09}). IR-luminous galaxies ($L_{\rm IR}>10^{10}$\,L$_{\odot}$) were discovered in large numbers by the \textit{IRAS} all sky survey in the 1980s (Soifer et al. 1984a\nocite{Soifer84a}; 1987a\nocite{Soifer87a}; 1987b\nocite{Soifer87b}; Sanders $\&$ Mirabel 1996\nocite{SM96}). It was noted that these sources are rare in the local Universe (e.g. Kim $\&$ Sanders 1998\nocite{KS98}) but much more numerous at earlier epochs (e.g. Takeuchi et al. 2005\nocite{Takeuchi05}), being responsible for about half the total light emitted from all galaxies integrated through cosmic time (e.g. Gispert et al. 2000\nocite{Gispert00}; Lagache et al. 2005\nocite{Lagache05}; Dole et al. 2006\nocite{Dole06}). Indeed the total star formation rate per unit volume (e.g. Hopkins $\&$ Beacom 2006\nocite{HB06}; Madau $\&$ Dickinson 2014\nocite{MD14}), at all epochs, is primarily made up of galaxies which are infrared-luminous (e.g. Takeuchi et al. 2005\nocite{Takeuchi05}).

Apart from high star-formation rates, IR-luminous galaxies are  also characterised by an AGN incidence rate which increases as a function of $L_{\rm IR}$, with the vast majority of the most luminous IR-galaxies at every epoch showing some kind of AGN signature (e.g. Goto 2005\nocite{Goto05}; Kartaltepe et al. 2010\nocite{Kartaltepe10}; Yuan et al. 2010\nocite{Yuan10}; Goto et al. 2011a\nocite{Goto11a}). Indeed, luminous QSOs are seen to be strong far-IR/submm emitters (e.g. Willott et al. 2000\nocite{Willott00}; Priddey $\&$ McMahon 2001\nocite{PM01}; Page et al. 2001\nocite{Page01}; 2004\nocite{Page04}; Priddey et al. 2007\nocite{Priddey07}; Tsai et al. 2015\nocite{Tsai15}; Podigachoski et al. 2015\nocite{Podigachoski15}; 2016\nocite{Podigachoski16}) and the plethora of extremely infrared-luminous sources recently discovered by the Wide-Field Infrared Survey Explorer (WISE; Wright et al. 2010\nocite{Wright10}) are thought to be primarily powered by AGN (e.g. Wu et al. 2012; Jones et al. 2014; Tsai et al. 2015\nocite{Tsai15}; Fan et al. 2016\nocite{Fan16}; Glikman et al. 2018\nocite{Glikman18}).

\begin{figure*}
\epsfig{file=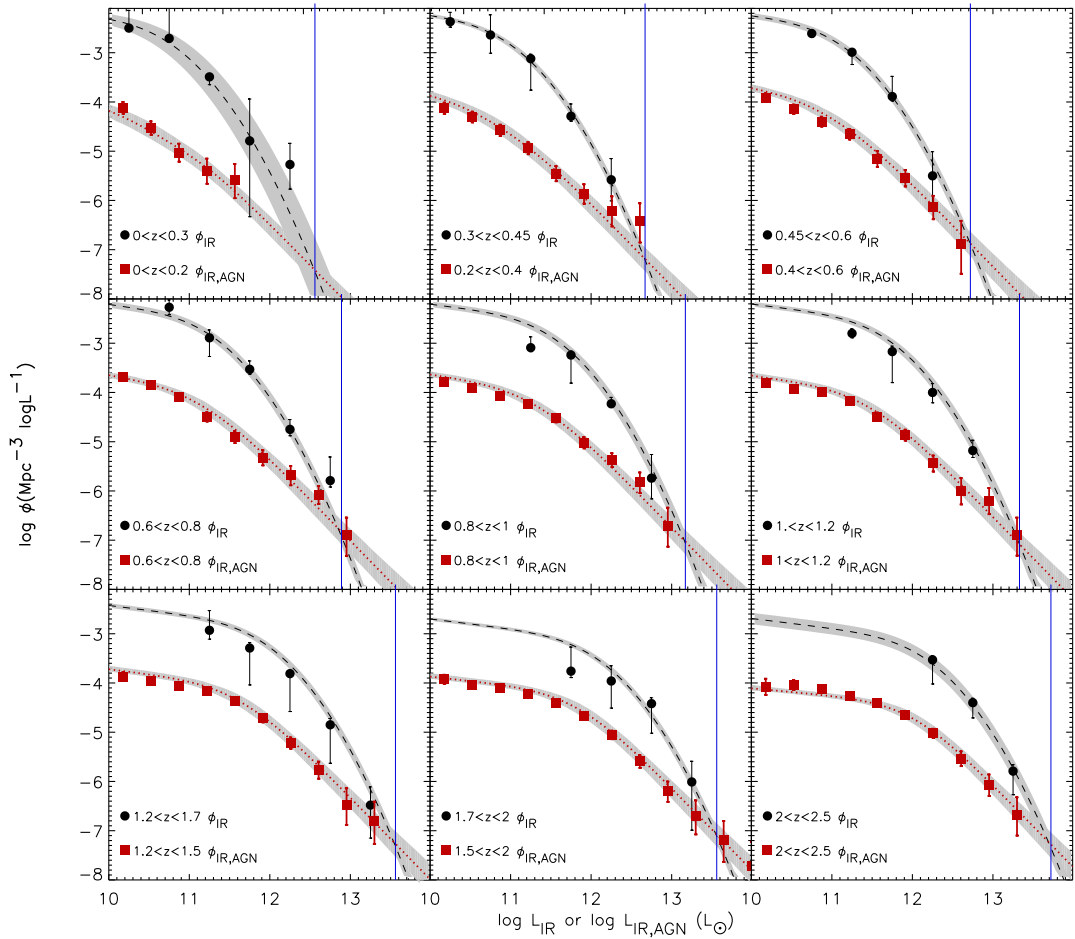,width=0.8\linewidth}\\
\caption{The IR LF ($\phi_{\rm IR}$) from Gruppioni et al. (2013): black filled circles. The corresponding functional form and 1$\sigma$ uncertainty are shown by the black dashed curve and shaded region. Shown with red squares is the IR AGN LF ($\phi_{\rm IR, AGN}$), derived from the hard X-ray LF in Aird et al. (2015). The red dotted curve and shaded outline represents the functional form and 1$\sigma$ uncertainty. The 9 panels correspond to different redshift bins as indicated. The vertical blue line indicates the luminosity where the parametric forms of the two LFs meet, $L_{\rm merge}$. The abscissa legend reads `log $L_{\rm IR}$ or log $L_{\rm IR, AGN}$' because $\phi_{\rm IR}$ is a function of $L_{\rm IR}$, whereas $\phi_{\rm IR, AGN}$ is a function of $L_{\rm IR, AGN}$. }
\label{fig:LFs}
\end{figure*}

The battle between stars and AGN in dust heating has been a topic of much contention going back as early as the 1990s (e.g. Gregorich et al. 1995\nocite{Gregorich95}; Genzel et al. 1998\nocite{Genzel98}; Soifer et al. 2000\nocite{Soifer00}; Klaas et al. 2001\nocite{Klaas01}; Davies et al. 2002\nocite{Davies02}; Franceschini et al. 2003\nocite{Franceschini03b}) and subsequently had a revival with the launch of the \textit{Herschel Space Observatory}\footnote{\textit{Herschel} is an ESA space observatory with science instruments provided by European-led Principal Investigator consortia and with important participation from NASA.} (Pilbratt et al. 2010\nocite{Pilbratt10}), designed to target the 70-500 $\mu$m wavelength range in which most of the Universe's obscured radiation emerges (e.g. Magnelli et al. 2010\nocite{Magnelli10}; Seymour et al. 2011\nocite{Seymour11}; Rovilos et al. 2012\nocite{Rovilos12}; Kirkpatrick et al. 2015\nocite{Kirkpatrick15}; Rawlings et al. 2015\nocite{Rawlings15}; Khan-Ali et al. 2015\nocite{Khan-Ali15}; Masoura et al. 2018\nocite{Masoura18} and many more). Recently, Symeonidis et al. (2016; hereafter S16\nocite{Symeonidis16}) and Symeonidis (2017; hereafter S17\nocite{Symeonidis17}) challenged the idea that far-IR emission is in all cases primarily powered by star-formation by showing that powerful AGN can dominate the entire infrared spectral energy distribution (SED). The implications of this are that the correlation between infrared luminosity and SFR must break down at high luminosities, at which point SFRs derived from infrared broadband photometry would be significantly overestimated. In order to gain further insight into the balance of power between AGN and stars in the IR-luminous galaxy population, Symeonidis $\&$ Page (2018; hereafter SP18\nocite{SP18}) and Symeonidis $\&$ Page (2019; hereafter SP19\nocite{SP19}) compared the behaviour of the IR galaxy LF to the IR AGN LF at $z\sim1-2$ and $z\sim0$ respectively. They discovered that at the high luminosity end of the AGN and galaxy LFs converge, suggesting that galaxies become AGN-dominated at high $L_{\rm IR}$. 

In this paper, we merge the work of SP18 and SP19 and subsequently develop it further with the following specific aims in mind: (i) to understand the shape of the IR LF in the $L_{\rm IR}\sim10^{10}-10^{15}$\,L$_{\odot}$ range, between $z=0$ and z$\sim$2.5, (ii) to separate the $L-z$ space into AGN-dominated and star-formation dominated regions, (iii) to examine the breakdown in the SFR-$L_{\rm IR}$ correlation and (iv) to quantify the effect of AGN in shaping the $L_{\rm IR}$ - dust temperature ($T_{\rm dust}$) relation. Our paper is structured as follows: in sections \ref{sec:method} and \ref{sec:results} we describe our method and results. The discussion and conclusions are presented in sections \ref{sec:discussion} and \ref{sec:conclusions}. Throughout, we adopt a concordance cosmology of H$_0$=70\,km\,s$^{-1}$Mpc$^{-1}$, $\Omega_{\rm M}$=1-$\Omega_{\rm \Lambda}$=0.3.

\begin{figure*}
\epsfig{file=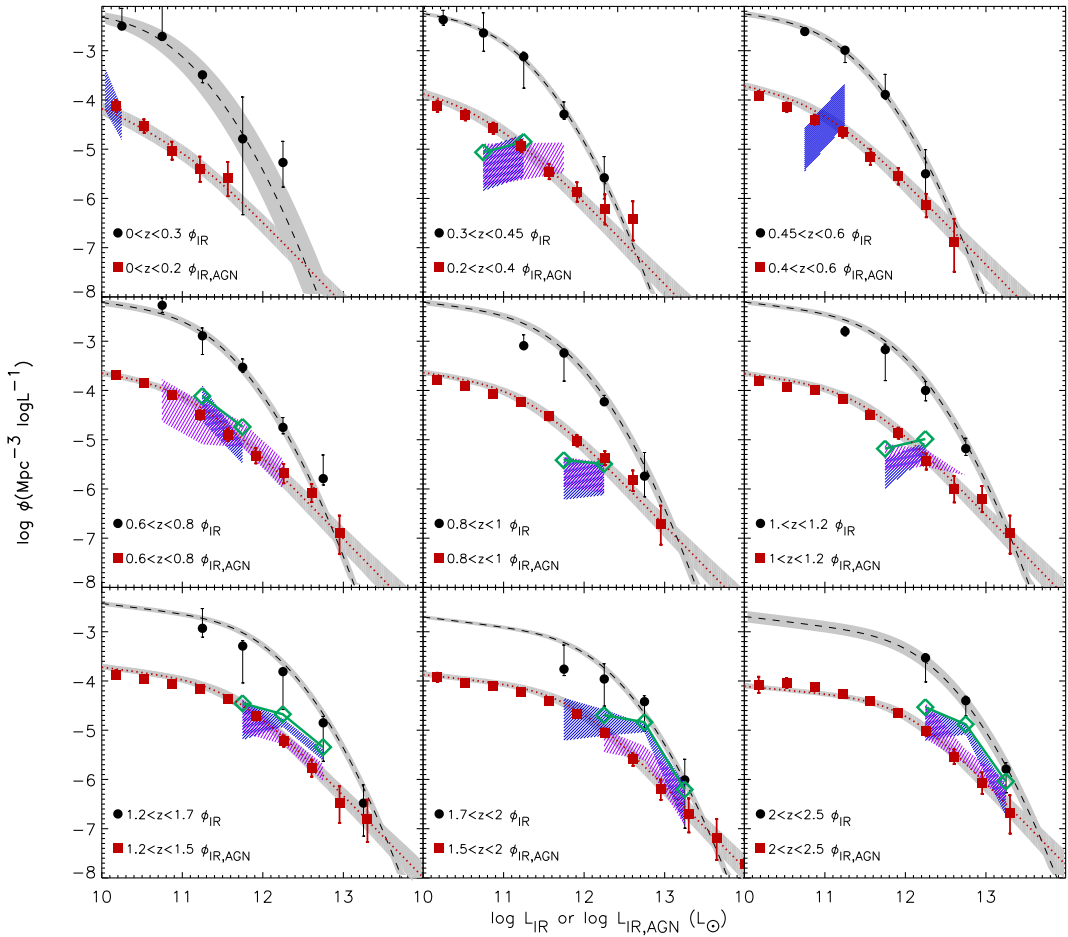,width=0.8\linewidth} \\
\caption{The IR LF ($\phi_{\rm IR}$) from Gruppioni et al. (2013): black filled circles. The corresponding functional form and 1$\sigma$ uncertainty are shown by the black dashed curve and shaded region. Shown with red squares is the IR AGN LF ($\phi_{\rm IR, AGN}$), derived from the hard X-ray LF in Aird et al. (2015). The red dotted curve and shaded outline represents the functional form and 1$\sigma$ uncertainty. The 9 panels correspond to different redshift bins as indicated. Included in this figure are also the IR space densities of AGN host galaxies as presented in Gruppioni et al. (2013). The space densities and corresponding uncertainties of sources fitted with type-1 AGN SEDs are indicated with the blue shaded area, whereas the sources fitted with type-2 AGN SEDs are indicated with the purple shaded area. Where both type-1 and type-2 data are present, the number densities are added and the total is indicated by the open green diamonds. The abscissa legend reads `log $L_{\rm IR}$ or log $L_{\rm IR, AGN}$' because $\phi_{\rm IR}$ is a function of $L_{\rm IR}$, whereas $\phi_{\rm IR, AGN}$ is a function of $L_{\rm IR, AGN}$.}
\label{fig:LFsG}
\end{figure*}

\begin{figure}
\epsfig{file=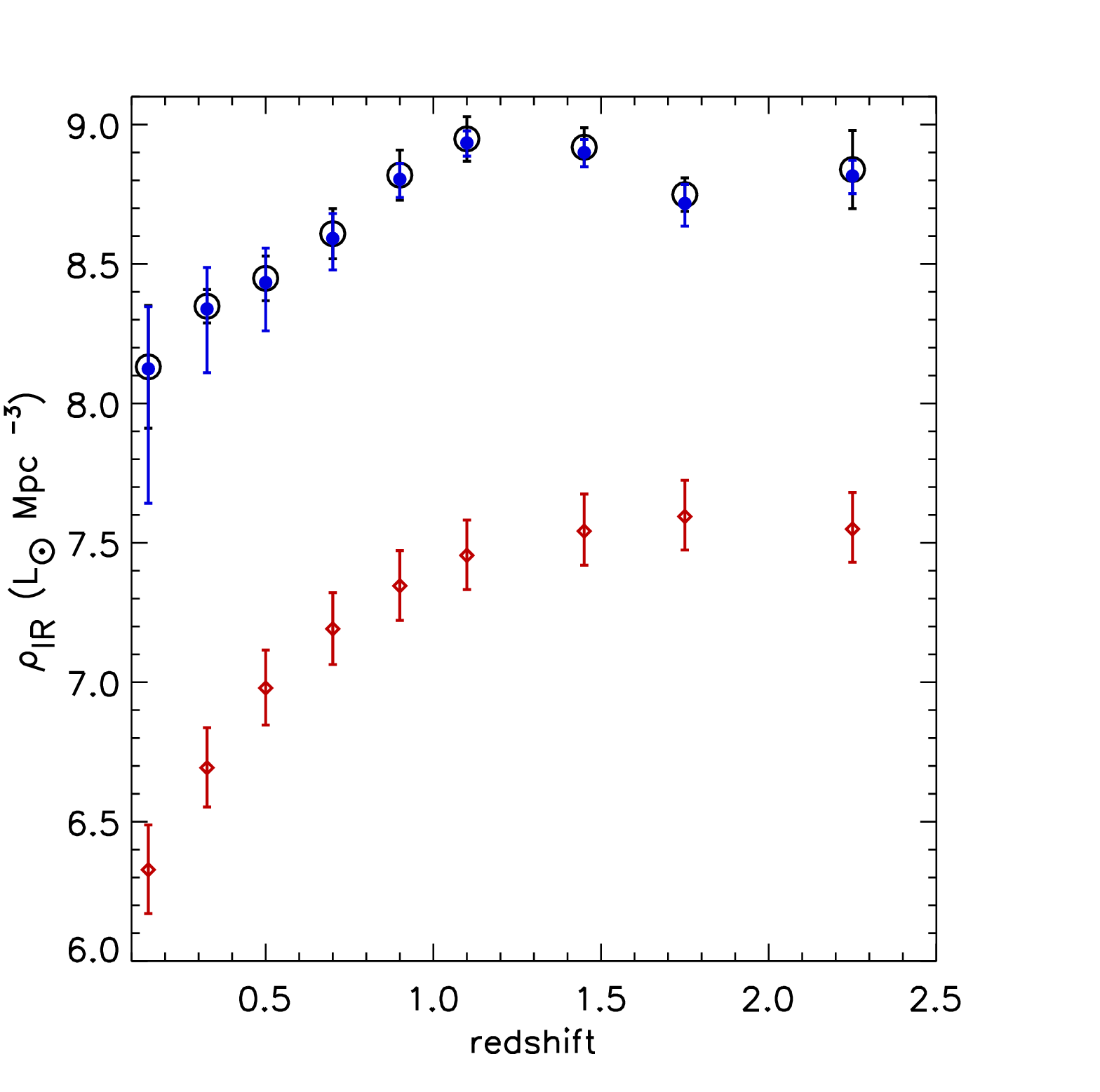,width=1.\linewidth} 
\caption{The luminosity density as a function of redshift: the total infrared luminosity density ($\rho_{\rm IR}$; large black open circles), the infrared luminosity density of AGN ($\rho_{\rm IR, AGN}$; red diamonds) and the infrared luminosity density of star-formation ($\rho_{\rm IR, SF}$; small blue filled circles). Table \ref{tab:lumden} lists the plotted data.}
\label{fig:lumdensity}
\end{figure}

\section{Method}
\label{sec:method}

We compare the galaxy LF and the AGN LF in the infrared (8--1000$\mu$m). This energy band is chosen for two main reasons:  (i) the IR LF is more complete than the UV/optical LFs at all redshifts as it includes galaxies which are heavily obscured in the UV/optical (ii) examining LFs in the 8--1000$\mu$m spectral range, rather than focusing on a particular monochromatic IR band, ensures that all IR-emitters are included irrespective of variations in their SEDs. 

As in SP18, for the IR galaxy LF ($\phi_{\rm IR}$) we use the one presented in Gruppioni et al. (2013\nocite{Gruppioni13}; hereafter G13). $\phi_{\rm IR}$ is a function of $L_{\rm IR}$, which includes the total dust-reprocessed emission from stars and AGN. The uncertainties on $\phi_{\rm IR}$ from G13 are a combination of Poisson errors and photometric redshift uncertainties derived through Monte Carlo simulations. The model fit to $\phi_{\rm IR}$ is the Saunders (1990\nocite{Saunders90}) function which behaves as a power-law for $L<L_{\star}$ and as a Gaussian for $L>L_{\star}$ (see G13 for more details). For the AGN LF we use the absorption-corrected hard X-ray (2-10\,keV) AGN LF from Aird et al. (2015\nocite{Aird15}; hereafter A15). The errors on $\phi_{\rm IR, AGN}$ are Poisson. The A15 AGN LF is fit with a double power-law model, whose parameters are themselves functions of redshift evaluated at the centre of the relevant bin (see A15 for more details). Note that we examine the behaviour of the luminosity functions up to $z\sim2.5$, because as discussed in G13, there is a severe lack of spectroscopic redshifts amongst the population that makes up the IR LF at $z>2.5$. 

We translate the X-ray AGN LF into an infrared AGN LF ($\phi_{\rm IR, AGN}$) as follows: first, hard X-ray luminosity is converted to optical luminosity at 5100$\AA$ ($\nu L_{\nu, 5100}$), adopting the equation from Maiolino et al. (2007\nocite{Maiolino07}), who derived it from the $\alpha_{\rm OX}$ relation reported in Steffen et al. (2006) by converting $L_{\rm 2 keV}$ in the Steffen et al. (2016) relation to $L_{\rm 2-10 keV}$ using $\Gamma=-1.7$. To be consistent with the work of A15 which assumes a $\Gamma$ of -1.9, we modify the Maiolino et al. (2007) equation by adding the constant C:  
\begin{equation}
\rm log\,[L_{2-10 keV}]=0.721\,\rm log\,[\nu L_{\nu} (5100\AA)]+ 11.78 + C
\end{equation}
where $C= \rm log(3.49) - log(4.14)$ and 3.49 is the value of $\frac{L_{2-10 keV}}{L_{2 keV}}$ for $\Gamma=-1.9$ whereas 4.14 is the value of $\frac{L_{2-10 keV}}{L_{2 keV}}$ for $\Gamma=-1.7$. A $\Gamma=-1.9$ is also favoured over $\Gamma=-1.7$ by studies of large samples of AGN X-ray spectra gathered by \textit{XMM-Newton} (e.g. Mateo et al. 2005; 2010; Page et al. 2006). 

Subsequently, to convert from $\nu L_{\nu, 5100}$ to infrared luminosity in the 8--1000$\mu$m range ($L_{\rm IR, AGN}$) we use the intrinsic AGN SED of S16, which represents the average optical-submm broadband emission from AGN. The $L_{\rm IR}$/$\nu L_{\nu, 5100}$ ratio for the S16 SED is 1.54. In section \ref{sec:otherseds} we also investigate effect of using other AGN SEDs with different $L_{\rm IR}$/$\nu L_{\nu, 5100}$ ratios.

Note that in this work we assume (i) the geometric unification of AGN, in which type-1 and type-2 AGN are intrinsically the same objects viewed from different angles and (ii) that both type-1 and type-2 AGN infrared luminosities scale in the same way with the accretion disc luminosity (e.g. Gandhi et al. 2009). Any differences in the shape of the intrinsic SED of type 2 and type 1 AGN are `washed out', since we only make use of the integrated 8--1000$\mu$m luminosity (e.g. Polletta 2006, 2007; Tsai et al. 2015). 

$\phi_{\rm IR, AGN}$ is now a function of $L_{\rm IR, AGN}$, not X-ray luminosity, where $L_{\rm IR, AGN}$ is the intrinsic IR luminosity of the AGN, i.e. it does not include the contribution of dust heated by starlight. Note that the A15 X-ray AGN LF does not include Compton-thick AGN. Therefore, following the prescription of A15, we scale the normalisation of the $\phi_{\rm IR, AGN}$ with their estimate of the Compton thick (CT) AGN fraction which is 34 per cent of the absorbed AGN population. In the A15 formulation the CT fraction is a constant fraction of the absorbed AGN population, but the absorbed AGN population fraction is itself a function of redshift and luminosity --- so indirectly the CT fraction is also a function of redshift and luminosity. 

At this stage, we take into account two forms of uncertainty in the conversion from the A15 X-ray LF to $\phi_{\rm IR, AGN}$: one related to the conversion from X-ray to optical luminosity and the other to the conversion from optical to infrared luminosity. For the former, we use the standard error on the mean $\alpha_{\rm OX}$ computed using the data in table 5 of Steffen et al. (2006), averaged over all bins. This corresponds to a 16.6 per cent (1$\sigma$) uncertainty on the $L_{\rm 2-10 keV}$/$\nu L_{\nu} (5100\AA)$ ratio. For the conversion from optical to infrared, we make use of the full set of individual intrinsic AGN SEDs used to derive the average S16 AGN SED (see S16 and S17), finding the (1$\sigma$) error on the mean $L_{\rm IR}$/$\nu L_{\nu} (5100\AA)$ ratio to be 9.4 per cent. Both of these are abscissa uncertainties, so we convert them to ordinate uncertainties on $\phi_{\rm IR, AGN}$ using the gradient of the luminosity function. The error on the $L_{\rm IR}$/$\nu L_{\nu} (5100\AA)$ ratio translates to a $\phi_{\rm IR, AGN}$ uncertainty in the range of 3--14 per cent for $L_{\rm IR, AGN}<$$10^{12}$\,L$_{\odot}$ and 14--15 per cent at $L_{\rm IR, AGN}>$$10^{12}$\,L$_{\odot}$, whereas the $L_{\rm 2-10 keV}$/$\nu L_{\nu} (5100\AA)$ ratio error translates to a $\phi_{\rm IR, AGN}$ uncertainty in the range of 3--14 per cent for $L_{\rm IR, AGN}<$$10^{12}$\,L$_{\odot}$ and 14--26 per cent for $L_{\rm IR, AGN}>$$10^{12}$\,L$_{\odot}$. These are added in quadrature to the A15 error on the functional form of $\phi_{\rm IR, AGN}$ in order to adjust the width of the $\phi_{\rm IR, AGN}$ 1$\sigma$ boundaries.

\section{Results}
\label{sec:results}

The data and functional forms of $\phi_{\rm IR, AGN}$ and $\phi_{\rm IR}$ are shown in Fig. \ref{fig:LFs} in 9 redshift bins within the $0<z<2.5$ interval. $\phi_{\rm IR}$ and $\phi_{\rm IR, AGN}$ are monotonically decreasing functions of $L_{\rm IR}$ and $L_{\rm IR, AGN}$ respectively (over the luminosity range considered here), and $\phi_{\rm IR} \geq \phi_{\rm IR, AGN}$. 
The $1.2<z<1.5$ ($z_{\rm centre}$=1.35) and $1.5<z<2$ ($z_{\rm centre}$=1.75) redshift bins are taken from SP18 and the remaining redshift bins are presented here for the first time. In the bins where the G13 and A15 results do not cover exactly the same redshift range, we also evaluated the parametric model of the A15 LF at the the centre of the G13 bins, finding the mean shift to be negligible at the bright end, so we use the original redshift bins for $\phi_{\rm IR, AGN}$ in Fig. \ref{fig:LFs}, as the AGN luminosity densities were calculated in those bins in A15.

Fig. \ref{fig:LFs} shows that at low luminosities, $\phi_{\rm IR}$ and $\phi_{\rm IR, AGN}$ are offset by up to 2\,dex, but this difference decreases with increasing luminosity, and eventually $\phi_{\rm IR}$ and $\phi_{\rm IR, AGN}$ converge. For the sake of consistency in all redshift bins, we define $L_{\rm merge}$ to be the luminosity at which the parametric forms of $\phi_{\rm IR, AGN}$ and $\phi_{\rm IR}$ meet. Note that although the data do not cover the $L_{\rm IR}\sim L_{\rm merge}$ parameter space in all bins, we are confident that $L_{\rm merge}$ is a good approximation of the luminosity of convergence of the two LFs. In some redshift bins one can see that the AGN and galaxy number densities are similar even before the parametric forms meet. Moreover, optical QSO surveys like the SDSS which cover large areas of sky to faint fluxes, have provided well-sampled AGN LFs to larger luminosities and smaller space densities than probed here. These show no change in the slope of the AGN LF down to space densities that are two orders of magnitude lower than probed by the A15 LF, and far beyond $L_{\rm merge}$ (e.g. Croom et al. 2009). 

\subsection{The number densities of AGN}
Although it is not possible to directly measure $L_{\rm IR, AGN}$ (AGN emission only) and corresponding number densities, one can measure the number densities of AGN as a function of $L_{\rm IR}$ (host+AGN emission). The latter measurements are then useful for comparing to a model of the former. In Fig \ref{fig:LFsG} we compare $\phi_{\rm IR, AGN}$ with the G13 AGN LF. G13 perform SED fitting on their sample of IR-selected galaxies and built the AGN LF by selecting only the sources that get flagged in the SED fitting process as hosting an AGN. The G13 AGN LF thus represents the space densities of candidate AGN hosts, and $L_{\rm IR}$ in this case is the emission from AGN and the host. On the other hand, in our work, $\phi_{\rm IR, AGN}$ represents the space densities of AGN as a function of $L_{\rm IR, AGN}$, i.e. emission from the AGN only. Fig \ref{fig:LFsG} shows that there is good agreement between the G13 AGN LF and our $\phi_{\rm IR}$.

\subsection{The infrared luminosity density}
By integrating $\phi_{\rm IR}$ and $\phi_{\rm IR, AGN}$ we calculate the total infrared luminosity density ($\rho_{\rm IR}$ and $\rho_{\rm IR, AGN}$ respectively) as a function of redshift. Subtracting $\rho_{\rm IR, AGN}$ from $\rho_{\rm IR}$ gives the IR luminosity density from star-formation ($\rho_{\rm IR, SF}$). These are plotted in Fig. \ref{fig:lumdensity}. The shape of $\rho_{\rm IR}$ and $\rho_{\rm IR, AGN}$ look similar, with an initial increase up to $z\sim1$ and a plateau thereafter. However, $\rho_{\rm IR}$ is a factor of 19--64 higher than $\rho_{\rm IR, AGN}$ and the contribution of AGN to the total infrared luminosity density ranges from $\sim$1.6 per cent at $z=$0.15 to $\sim$5 per cent at $z= 2.25$ (see table \ref{tab:lumden}). The change in fractional AGN contribution with redshift is only significant at the $<$2$\sigma$ level, thus there is no evidence that the contribution of AGN to the total IR energy budget is dependent on redshift. 

Our results indicate that current estimates of the cosmic SFR density are only marginally affected by AGN contamination. This is slightly different to what is found by Gruppioni et al. (2015\nocite{Gruppioni15}), who show a small but significant AGN contribution particularly at intermediate redshift (z$\sim$2-2.5). We believe this difference is down to the different approaches in estimating the AGN contribution. Gruppioni et al. (2015) use the Delvecchio et al. (2014\nocite{Delvecchio14}) results who computed the AGN contribution on an object per object basis using multi-component SED fitting: they find 37 per cent AGN incidence in their IR-selected sample. This is higher than what is reported with traditional AGN indicators  --- for example Symeonidis et al. (2014\nocite{Symeonidis14b}) find $\sim$20 per cent AGN incidence in a sample of IR-selected galaxies when examining the hardness ratio, mid-IR colours, optical and X-ray variability, radio loudness and high excitation optical lines. We believe the difference in the AGN incidence rate is because multi-component SED fitting results in a larger fraction of sources requiring some level of AGN contribution to their total infrared luminosity. 

In any case both the Gruppioni et al. (2015) results and the ones presented here consistently indicate that AGN are not the primary contributors to $\rho_{\rm IR}$ at any redshift.

\begin{table}
\centering
\caption{$\rho_{\rm IR}$, $\rho_{\rm IR, AGN}$ and $\rho_{\rm IR, SF}$ (in log[L$_{\odot}\rm Mpc^{-3}$]) as shown in Fig. \ref{fig:lumdensity}. The log 1$\sigma$ lower and upper values of $\rho_{\rm IR}$, $\rho_{\rm IR, AGN}$ and $\rho_{\rm IR, SF}$ are also listed. The last column shows the fractional contribution of AGN to $\rho_{\rm IR}$ as a percentage. The redshifts are quoted at the middle of the bins.}
\begin{tabular}{|l|c|c|c|c|}
\hline 
z  & log\,$\rho_{\rm IR}$ & log\,$\rho_{\rm IR, AGN}$& log\,$\rho_{\rm IR, SF}$ & $\rho_{\rm IR, AGN}$/$\rho_{\rm IR} (\%)$ \\
\hline
             0.15&  8.13$_{7.91}^{8.35}$&  6.33$_{6.17}^{6.49}$& 8.12$_{7.64}^{8.35}$&1.57\\
      0.325&  8.35$_{8.29}^{8.41}$&  6.69$_{6.55}^{6.84}$& 8.34$_{8.27}^{8.40}$&2.21\\
      0.5&  8.45$_{8.37}^{8.53}$&  6.98$_{6.85}^{7.12}$& 8.43$_{8.33}^{8.52}$&3.40\\
      0.7&  8.61$_{8.52}^{8.70}$&  7.19$_{7.06}^{7.32}$& 8.59$_{8.47}^{8.69}$&3.83\\
      0.9&  8.82$_{8.73}^{8.91}$&  7.35$_{7.22}^{7.47}$& 8.80$_{8.69}^{8.90}$&3.37\\
       1.1&  8.95$_{8.87}^{9.03}$&  7.46$_{7.33}^{7.58}$& 8.93$_{8.83}^{9.02}$&3.21\\
       1.45&  8.92$_{8.85}^{8.99}$&  7.54$_{7.42}^{7.68}$& 8.90$_{8.81}^{8.97}$&4.20\\
       1.75&  8.75$_{8.69}^{8.81}$&  7.59$_{7.47}^{7.72}$& 8.72$_{8.64}^{8.78}$&7.01\\
       2.25&  8.84$_{8.70}^{8.98}$&  7.55$_{7.43}^{7.68}$& 8.82$_{8.59}^{8.96}$&5.14\\
\hline
\end{tabular}
\label{tab:lumden}
\end{table}

\begin{figure*}
\epsfig{file=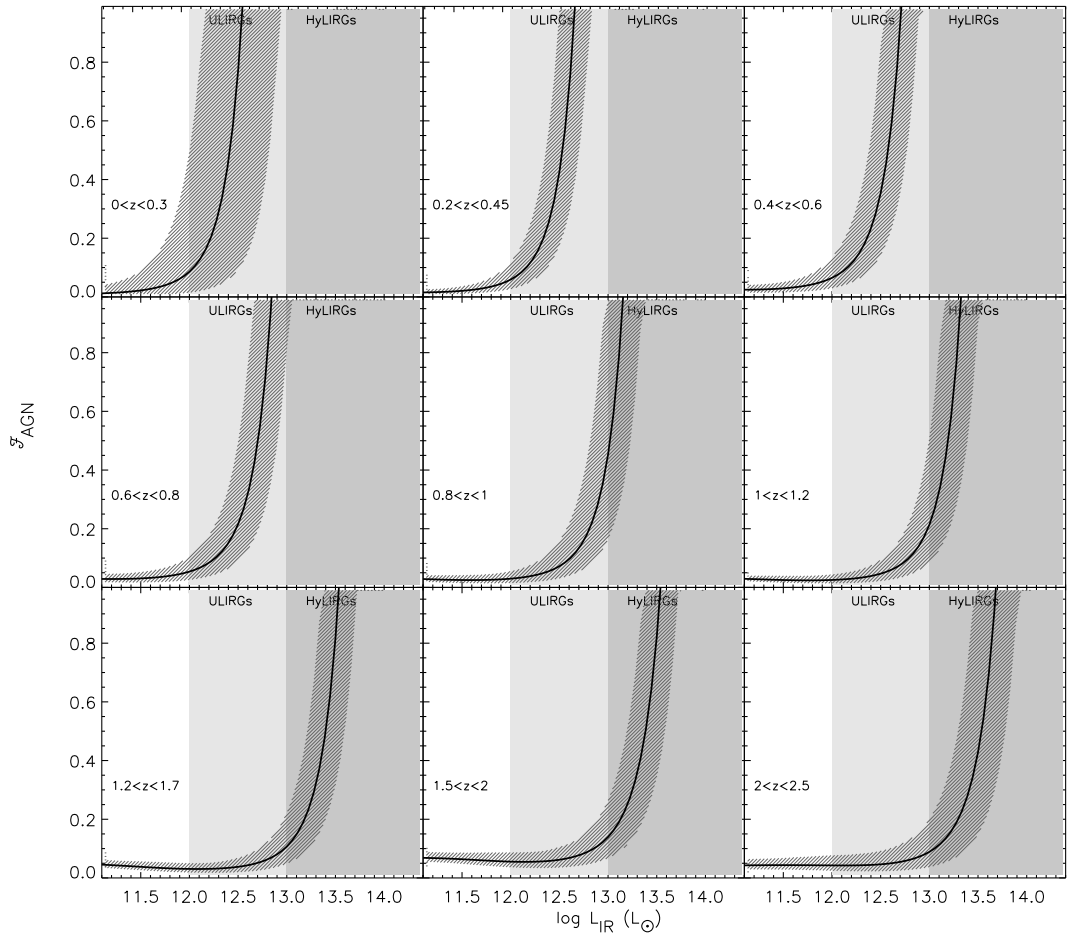,width=0.8\linewidth} 
\caption{Plotted here is the ratio of $\phi_{\rm IR, AGN}$ to $\phi_{\rm IR}$ ($\mathcal F_{\rm AGN}$) which provides a simple estimate of the fraction of AGN-dominated sources as a function of $L_{\rm IR}$. The shaded outline to the curve represents the 1$\sigma$ uncertainty interval calculated from the uncertainties of the LFs in Fig. \ref{fig:LFs}. The panels correspond to different redshift bins as indicated. The shaded vertical bands represent the ULIRG ($10^{12}<L_{\rm IR}<10^{13}$\,L$_{\odot}$) and HyLIRG ($L_{\rm IR}>10^{13}$\,L$_{\odot}$) regimes.}
\label{fig:ratio}
\end{figure*}

\begin{figure}
\epsfig{file=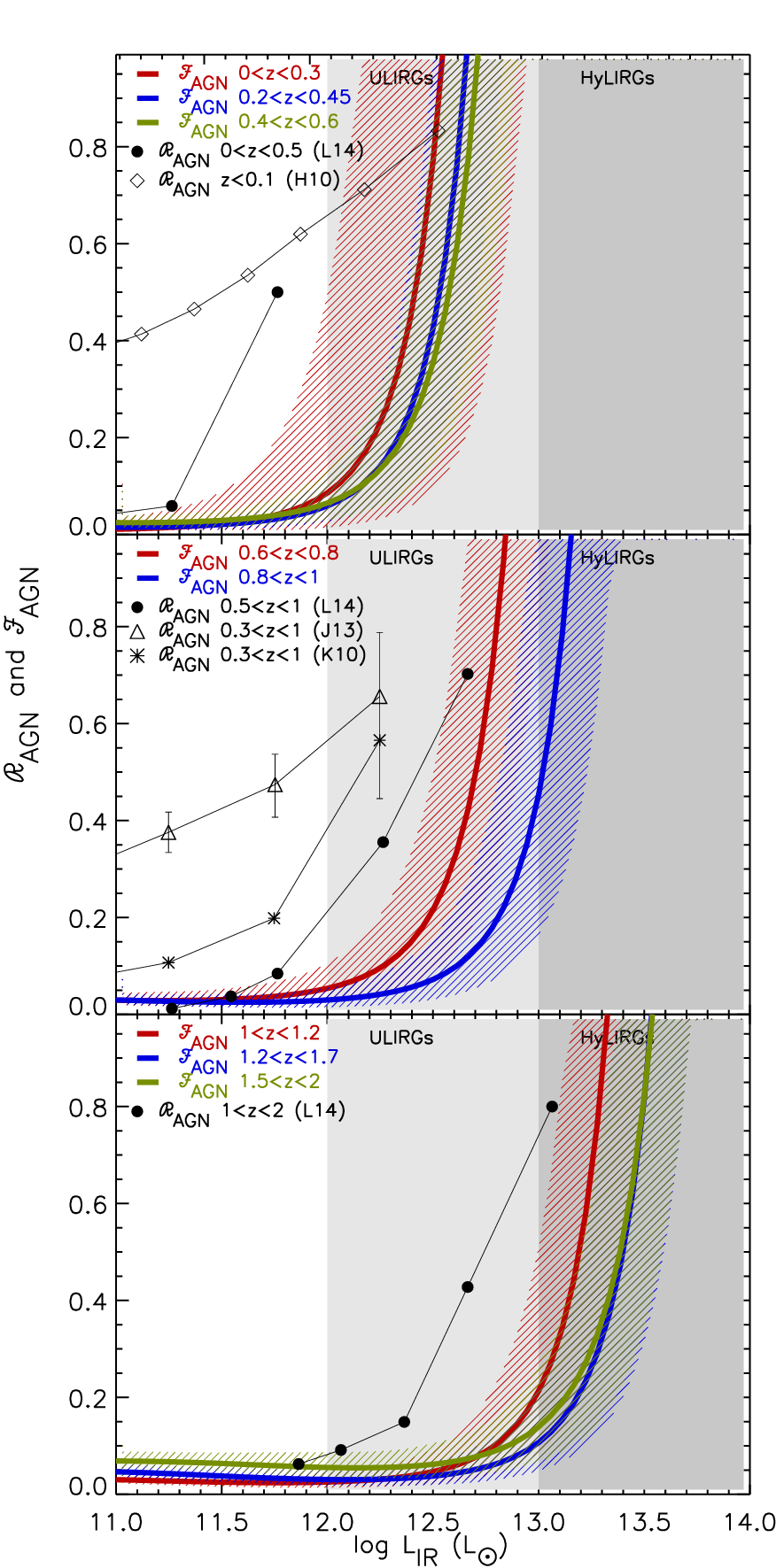,width=0.9\linewidth} 
\caption{The ratio of $\phi_{\rm IR, AGN}$ to $\phi_{\rm IR}$ ($\mathcal F_{\rm AGN}$) in the $0<z<0.6$ range (top pannel), $0.6<z<1$ range (middle panel) and $1<z<2$ (bottom panel). 
Also shown is the AGN incidence rate, $\mathcal R_{\rm AGN}$, reported in Hwang et al. (2010; H10), Kartaltepe et al. (2010; K10), Juneau et al. (2013; J13) and Lemaux et al. (2014; L14) compared to $\mathcal F_{\rm AGN}$ in similar redshift ranges. }
\label{fig:ratio_AGNfrac}
\end{figure}

\subsection{Defining the AGN fraction}
\label{sec:AGNfrac}
The ratio of $\phi_{\rm IR, AGN}$ to $\phi_{\rm IR}$ (hereafter referred to as $\mathcal F_{\rm AGN}$) provides a simple estimate of the fraction of AGN-dominated sources as a function of $L_{\rm IR}$ (see also SP18 and SP19). This definition assumes that galaxies are either entirely AGN-powered or star-formation-powered (i.e. there is no mixing) and $\mathcal F_{\rm AGN}$ essentially represents $n_{\rm AGN}/(n_{\rm AGN}+n_{\rm SF}$), where $n_{\rm AGN}$ is the number of AGN-powered galaxies and $n_{\rm SF}$ is the number of star-formation-powered galaxies. Although this definition has its limitations at low luminosities where there might be substantial mixing between emission from stars and AGN, as we approach the high luminosity regime where the luminosity functions start converging, $L_{\rm IR} \sim L_{\rm IR, AGN}$, i.e. the AGN infrared emission dominates the $L_{\rm IR}$. As a result, we expect that $\mathcal F_{\rm AGN}$ adequately traces the AGN-dominated fraction of galaxies at least in the high luminosity regime. 

Fig. \ref{fig:ratio} shows $\mathcal F_{\rm AGN}$ as a function of $L_{\rm IR}$, calculated by dividing the $\phi_{\rm IR, AGN}$ parametric model by the $\phi_{\rm IR}$ parametric model in each redshift bin. Note that at all redshifts, the contribution of the AGN to the total infrared luminosity and hence the fraction of AGN-dominated sources is small at low $L_{\rm IR}$, but undergoes a rapid increase with increasing $L_{\rm IR}$, and at high $L_{\rm IR}$, the population becomes AGN dominated. Note that the whole curve shifts rightwards with increasing redshift, suggesting that the luminosity at which the infrared galaxy population becomes AGN-dominated increases as a function of redshift.

In Fig. \ref{fig:ratio_AGNfrac} we compare $\mathcal F_{\rm AGN}$ with the AGN incidence rate ($\mathcal R_{\rm AGN}$) as reported in Hwang et al. (2010\nocite{Hwang10}), Kartaltepe et al. (2010\nocite{Kartaltepe10}), Juneau et al. (2013\nocite{Juneau13}) and Lemaux et al. (2014\nocite{Lemaux14}), aiming to compare the same redshift ranges as much as possible. The different relations shown by these works are likely a result of the AGN selection criteria in the samples used. It is interesting to note that at a given $L_{\rm IR}$, the fraction of galaxies hosting AGN is much higher than the fraction of AGN-dominated galaxies, so the increase in the latter is very easily accommodated by the increase in the former. This suggests that at $L_{\rm merge}$, almost all galaxies host AGN. This is consistent with what G13 also find, namely that the sources that make up the high luminosity tail of $\phi_{\rm IR}$ are consistently fitted with SED models that have a strong AGN component. 

\subsection{The evolution of the AGN fraction with redshift}
\begin{figure}
\epsfig{file=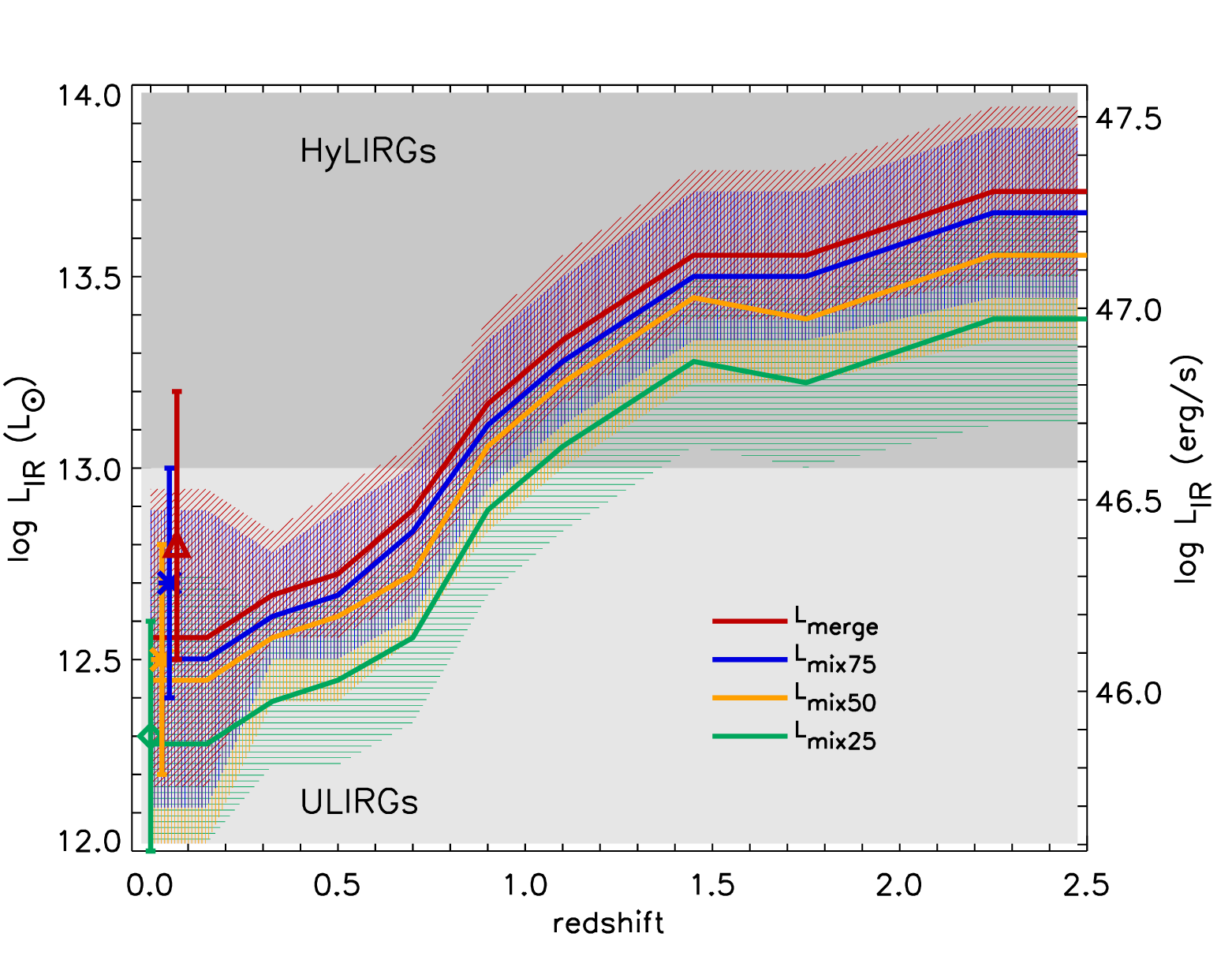,width=1.1\linewidth} 
\caption{The evolution of $L_{\rm mix25}$ (green), $L_{\rm mix50}$ (yellow), $L_{\rm mix75}$ (blue) and $L_{\rm merge}$ (red) representing the fraction of AGN-dominated sources at the 25, 50, 70 and 100 per cent levels respectively. The corresponding hatched regions represent the 1$\sigma$ uncertainties on these quantities derived from the uncertainties in $\mathcal F_{\rm AGN}$ as shown in Fig \ref{fig:ratio}. The symbols at $z<0.1$ represent the values for the local Universe taken from Symeonidis $\&$ Page (2019): green diamond for $L_{\rm mix25}$, yellow asterisk for $L_{\rm mix50}$, blue asterisk for $L_{\rm mix75}$ and red triangle for $L_{\rm merge}$. They are slightly offset in redshift for more clarity. }
\label{fig:breaklum}
\end{figure}

Earlier we defined $L_{\rm merge}$ as the luminosity at which $\phi_{\rm IR, AGN}$ and $\phi_{\rm IR}$ meet and hence $\mathcal F_{\rm AGN}$=1. We now also define the \textit{mixing luminosity} at $\mathcal F_{\rm AGN}$=0.25 ($L_{\rm mix25}$), $\mathcal F_{\rm AGN}$=0.5 ($L_{\rm mix50}$) and $\mathcal F_{\rm AGN}$=0.75 ($L_{\rm mix75}$) to be where the fraction of AGN-dominated sources is 25, 50 and 75 per cent respectively. Fig. \ref{fig:breaklum} shows these quantities as a function of redshift, as well as the measurements for the local ($z<0.1$) Universe from SP19. The SP19 values 
of $L_{\rm mix25}$, $L_{\rm mix50}$, $L_{\rm mix75}$ and $L_{\rm merge}$ at $z<0.1$ do not exactly match the estimates of these quantities at $0<z<0.3$ derived in the current work, although they are entirely consistent within the errors. The reasons for this difference are the redshift ranges probed ($z<0.1$ in SP19 versus $0<z<0.3$ in this work) as well as the LFs used --- SP19 used the X-ray LFs from Sazonov et al. (2007), Tueller et al. (2008) and Ueda et al. (2011) and the IR LFs from Goto et al. (2011) and Saunders et al. (1990), whereas in this work we use the A15 and G13 LFs. 

$L_{\rm mix25}$, $L_{\rm mix50}$, $L_{\rm mix75}$ and $L_{\rm merge}$ increase with redshift, not surprising as both the AGN and IR LFs undergo redshift evolution (Fig \ref{fig:LFs}). It is interesting to note that at a given $L_{\rm IR}$ the fraction of AGN-dominated sources is higher at low redshift than it is at high redshift.

\begin{figure}
\epsfig{file=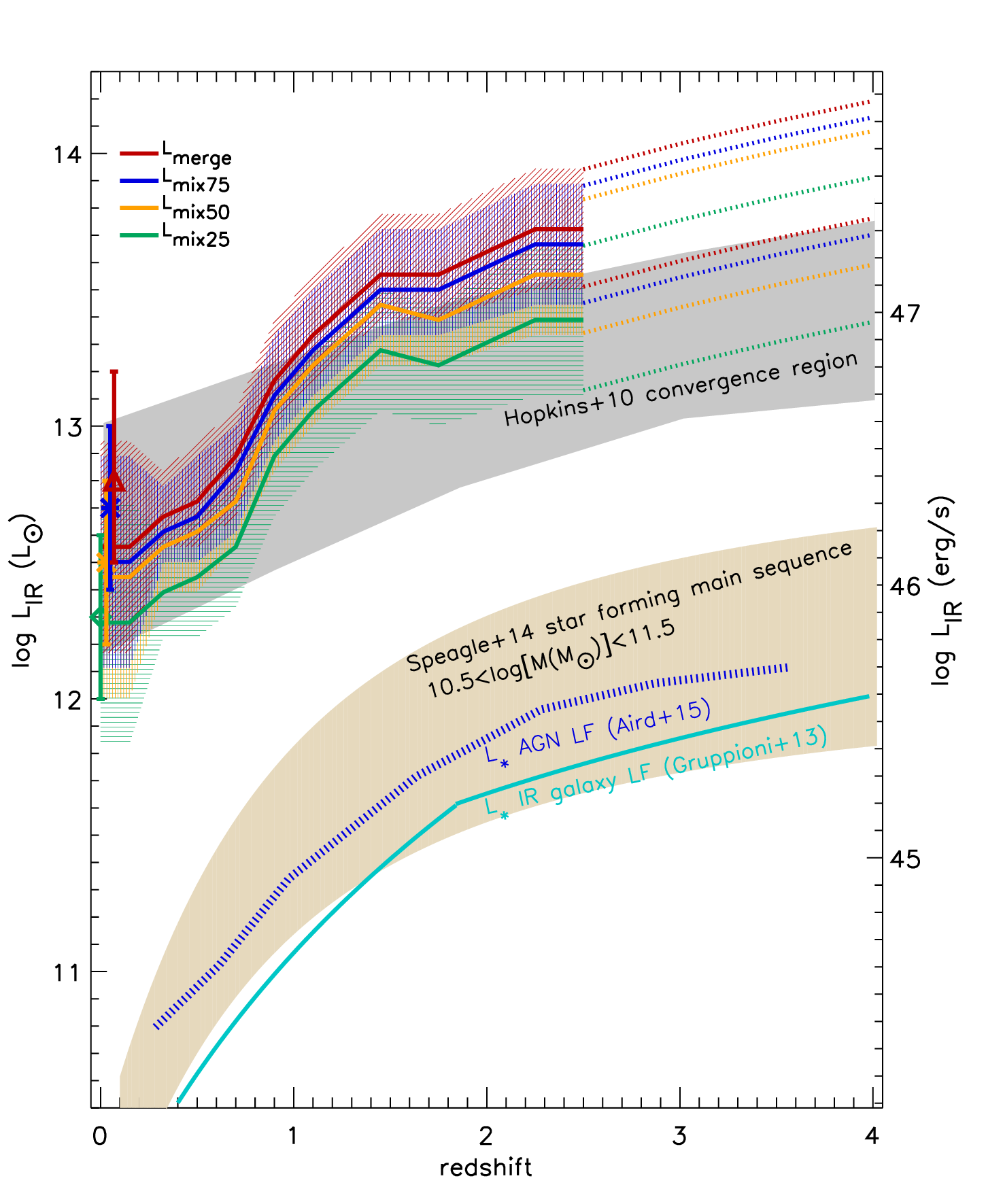,width=1.1\linewidth} 
\caption{The evolution of $L_{\rm mix25}$ (green), $L_{\rm mix50}$ (yellow), $L_{\rm mix75}$ (blue) and $L_{\rm merge}$ (red) representing the fraction of AGN-dominated sources at the 25, 50, 75 and 100 per cent levels respectively. The corresponding hatched regions represent the 1$\sigma$ uncertainties on these quantities derived from the uncertainties in $\mathcal F_{\rm AGN}$ as shown in Fig \ref{fig:ratio}. The dotted lines represent the extrapolated $L_{\rm mix25}$, $L_{\rm mix50}$, $L_{\rm mix75}$ and $L_{\rm merge}$ 1$\sigma$ boundaries up to $z\sim4$, computed by evolving the luminosity by $(1+z)^{1.62}$. 
The symbols at $z<0.1$ represent the values for the local Universe taken from Symeonidis $\&$ Page (2019): green diamond for $L_{\rm mix25}$, yellow asterisk for $L_{\rm mix50}$, blue asterisk for $L_{\rm mix75}$ and red triangle for $L_{\rm merge}$. They are slightly offset in redshift for clarity. The grey shaded region is taken from Hopkins et al. (2010) and indicates their modelled convergence region where objects change from star-formation to AGN dominated. Also shown are the $L_{\star}$ from the Gruppioni et al. (2013) IR LF (solid turquoise curve) and the $L_{\star}$ from the AGN LF of A15 converted to the IR (vertical dashed blue curve). Finally, the evolution of the SFR-M$_{\star}$ relation with redshift, evaluated at M$_{\star}$=10$^{10.5}$\,M$_{\odot}$ and M$_{\star}$ =10$^{11.5}$\,M$_{\odot}$, taken from Speagle et al. (2014), is plotted as a pale brown region. }
\label{fig:breaklum_extrapolate}
\end{figure}

\begin{figure*}
\epsfig{file=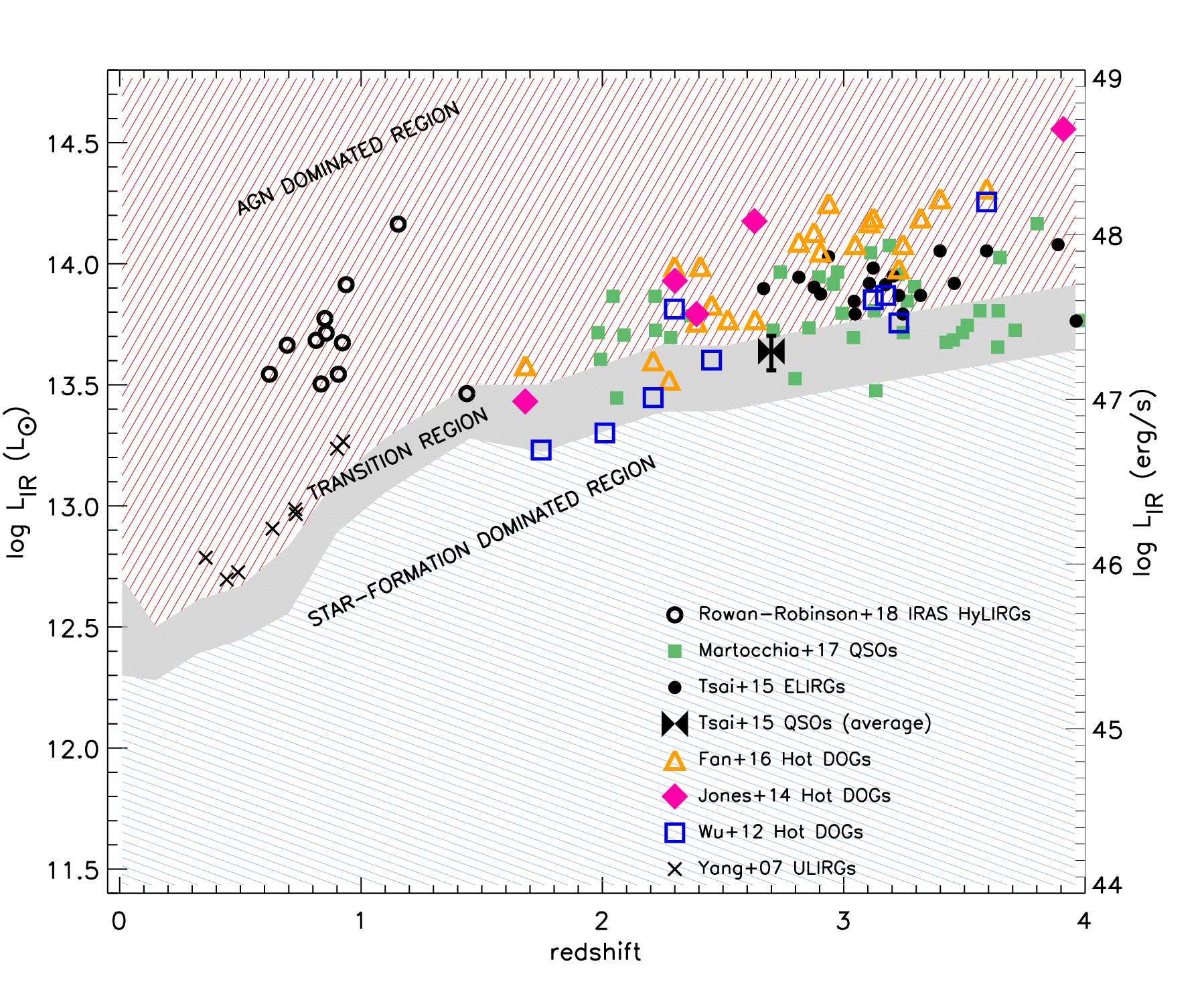,width=0.85\linewidth} 
\caption{Partitioning the $L_{\rm IR}-z$ space into a star-formation-dominated, a transition and an AGN-dominated region. The transition region is within $L_{\rm mix25}$ and $L_{\rm mix75}$ (see Fig. \ref{fig:breaklum_extrapolate}). The AGN-dominated region is defined as being above $L_{\rm mix75}$. Overplotted are various samples from the literature most of which are claimed to be amongst the most luminous at each redshift.}
\label{fig:breaklum_sources}
\end{figure*}

In Fig. \ref{fig:breaklum_extrapolate} we compare our results with the convergence region modelled by Hopkins et al. (2010), defined as the locus of convergence between the galaxy and AGN LF, with the width of this region representing the uncertainty in the convergence point. Note that this is equivalent to our definition of $L_{\rm merge}$ and its corresponding uncertainties. Hopkins et al. derive their boundaries theoretically, using a semi-empirical approach, starting with a halo occupation model convolved with observables such as the stellar mass function and then evolved using the prescriptions from hydrodynamical simulations for the distribution of SFRs and $L_{\rm IR}$ in obscured AGN, quiescent galaxies and merger-induced starbursts in order to construct LFs. They assume that it is only obscured AGN that make a significant contribution to the infrared, stating that only up to 5 per cent of the bolometric luminosity of unobscured AGN is emitted in the far-IR. This fraction is consistent with what was proposed in S17, although the latter study showed that that it also applies to unobscured AGN. It is interesting to note that the convergence region in the Hopkins et al. formulation is in broad agreement with our work, almost completely overlapping with $L_{\rm merge}$ until about $z\sim0.8$. There is less pronounced overlap thereafter, however it has been noted that hydrodynamical simulations and semi-analytic models, often underestimate the high-luminosity end of the IR LF and the high-mass end of the mass function at high redshift (e.g. Gruppioni et al. 2015). As a result, it is possible that the Hopkins et al. approach might be underestimating the convergence region with increasing redshift. 

In Fig. \ref{fig:breaklum_extrapolate} we also show the Speagle et al. (2014) locus of the `main sequence of star-formation' (SFR - M$_{\star}$ relation) evaluated in the log $[M_{\star}(\rm M_{\odot})]=10.5-11.5$ range, using the Kennicutt (1998) calibration to convert SFR to $L_{\rm IR}$.
In addition we plot the $L_{\star}$ of $\phi_{\rm IR}$ from G13, and the $L_{\star}$ of the $\phi_{\rm IR, AGN}$ from A15 converted to the IR, although note that the two $L_{\star}$ functions are not directly comparable because the two LFs are fitted with different parametric forms. As expected the region described by $L_{\rm mix25}$, $L_{\rm mix50}$, $L_{\rm mix75}$ and $L_{\rm merge}$ is offset from the `knee' ($L_{\star}$) of $\phi_{\rm IR}$ and $\phi_{\rm IR, AGN}$. $L_{\rm mix25}$ is about 1.5-2 dex higher than the G13 $L_{\star}$ ($L_{\rm mix25} \sim 70 L_{\star}$) and $L_{\rm merge}$ is offset by $\sim 2.5$\,dex ($L_{\rm merge} \sim 120 L_{\star}$). The large offset from $L_{\star}$ and the `star-forming sequence' locus further illustrates the point that the AGN contribution to the total emission is not significant for the bulk of the star-forming galaxy population. 

Note that our computed $L_{\rm mix25}$, $L_{\rm mix50}$, $L_{\rm mix75}$ and $L_{\rm merge}$ do not extend past $z\sim2.5$. However, as they are all derived by dividing the luminosity functions, they are linked to how the luminosity functions themselves evolve and hence it is reasonable to assume that they evolve in a similar fashion to $L_{\star}$. We thus extrapolate $L_{\rm mix25}$, $L_{\rm mix50}$, $L_{\rm mix75}$ and $L_{\rm merge}$ by evolving them in the same way as the G13 $L_{\star}$, namely $L \propto (1+z)^{1.62}$ for $z>1.85$. The extrapolated quantities are shown as dotted lines in Fig. \ref{fig:breaklum_extrapolate}.

\begin{table}
\centering
\caption{Data for Fig. \ref{fig:breaklum}, indicating the values of $L_{\rm mix25}$, $L_{\rm mix50}$, $L_{\rm mix75}$ and $L_{\rm merge}$ at the middle of the redshift bins shown in Fig. \ref{fig:ratio}. Also included are the extrapolated $L_{\rm mix25}$, $L_{\rm mix50}$, $L_{\rm mix75}$ and $L_{\rm merge}$ shown in Fig \ref{fig:breaklum_extrapolate}. The log 1$\sigma$ upper and lower values of $L_{\rm mix25}$, $L_{\rm mix50}$, $L_{\rm mix75}$ and $L_{\rm merge}$ are also quoted. }
\begin{tabular}{| l |  c | c  |c  |c   |}
\hline 
redshift  &  log\,$L_{\rm mix25}$ & log\,$L_{\rm mix50}$ & log\,$L_{\rm mix75}$& log\,$L_{\rm merge}$\\
& (L$_{\odot}$) & (L$_{\odot}$) &(L$_{\odot}$)& (L$_{\odot}$) \\
\hline
          $<0.1^a$&12.30$_{12.00}^{12.60}$ &12.50$_{12.20}^{12.80}$ &12.70$_{12.40}^{13.00}$ &12.80$_{12.50}^{13.20}$\\
      0.15&12.28$_{11.84}^{12.72}$ &12.45$_{12.00}^{12.83}$ &12.50$_{12.11}^{12.89}$ &12.56$_{12.17}^{12.95}$\\
      0.325&12.39$_{12.22}^{12.61}$ &12.56$_{12.39}^{12.72}$ &12.61$_{12.50}^{12.78}$ &12.67$_{12.56}^{12.83}$\\
      0.5&12.45$_{12.22}^{12.67}$ &12.61$_{12.39}^{12.78}$ &12.67$_{12.50}^{12.89}$ &12.72$_{12.56}^{12.95}$\\
      0.7&12.56$_{12.34}^{12.78}$ &12.72$_{12.56}^{12.95}$ &12.83$_{12.61}^{13.00}$ &12.89$_{12.67}^{13.06}$\\
      0.9&12.89$_{12.67}^{13.11}$ &13.06$_{12.83}^{13.28}$ &13.11$_{12.95}^{13.33}$ &13.17$_{13.00}^{13.39}$\\
       1.1&13.06$_{12.83}^{13.28}$ &13.22$_{13.00}^{13.44}$ &13.28$_{13.11}^{13.50}$ &13.33$_{13.17}^{13.56}$\\
       1.45$^b$&13.28$_{13.06}^{13.50}$ &13.44$_{13.22}^{13.61}$ &13.50$_{13.33}^{13.72}$ &13.56$_{13.39}^{13.78}$\\
       1.75$^b$&13.22$_{13.00}^{13.44}$ &13.39$_{13.22}^{13.61}$ &13.50$_{13.33}^{13.72}$ &13.56$_{13.39}^{13.78}$\\
       2.25&13.39$_{13.11}^{13.67}$ &13.56$_{13.33}^{13.83}$ &13.67$_{13.44}^{13.89}$ &13.72$_{13.50}^{13.94}$\\
       2.5$^c$&13.39$_{13.13}^{13.66}$ &13.55$_{13.34}^{13.83}$ &13.66$_{13.45}^{13.88}$ &13.72$_{13.51}^{13.94}$\\
       3.0$^c$&13.49$_{13.23}^{13.76}$ &13.65$_{13.44}^{13.93}$ &13.76$_{13.55}^{13.98}$ &13.82$_{13.61}^{14.04}$\\
       3.5$^c$&13.57$_{13.31}^{13.84}$ &13.73$_{13.52}^{14.01}$ &13.84$_{13.63}^{14.06}$ &13.90$_{13.69}^{14.12}$\\
       4.0$^c$&13.64$_{13.38}^{13.91}$ &13.80$_{13.59}^{14.08}$ &13.91$_{13.70}^{14.13}$ &13.97$_{13.76}^{14.19}$\\     
\hline
\multicolumn{5}{l}{\textit{Notes:}}\\
\multicolumn{5}{l}{a: data from SP19}\\
\multicolumn{5}{l}{b: redshifts bins first explored in SP18 but data are from this work} \\
\multicolumn{5}{l}{c: extrapolation (see Fig \ref{fig:breaklum_extrapolate})} \\
\hline
\end{tabular}
\label{table:measurements}
\end{table}

\subsection{Partitioning the $L_{\rm IR}-z$ space}

Based on our results, we create a diagnostic diagram which serves to separate the $L-z$ space into a star-formation-dominated, a transition and an AGN-dominated region (Fig \ref{fig:breaklum_sources}). We define the AGN-dominated region as starting from $L_{\rm mix75}$ (Fig. \ref{fig:breaklum_extrapolate}), the transition region to be between $L_{\rm mix25}$ and $L_{\rm mix75}$ and the star formation dominated region at $L_{\rm IR}<L_{\rm mix25}$. Note that this diagram is not designed for classifying individual galaxies as AGN-dominated or star-formation-dominated, rather it reflects the dominance of populations in $L-z$ space. It is thus perfectly plausible that some sources in the star formation dominated region will be AGN-dominated in the IR. However, the more luminous a galaxy is the more likely it is that it will be AGN-dominated, and above a certain luminosity, it becomes a reasonable expectation that \textit{individual} galaxies can be assumed to be entirely AGN-powered.

We populate Fig \ref{fig:breaklum_sources} with various samples from the literature, selected to be (amongst) the most luminous at the redshifts probed.  Fig \ref{fig:breaklum_sources} includes optically unobscured QSOs from Tsai et al. (2015; see also S17), intermediate redshift ULIRGs from Yang et al. (2007\nocite{Yang07}) and the \textit{IRAS}-selected HyLIRGs from Rowan-Robinson et al. (2018\nocite{RR18}). We also plot sources from the WISSH project (Bischetti et al. 2007) which includes WISE $f_{22}>3$\,mJy sources with SDSS counterparts at $z>1.5$ (Martocchia et al. 2017). Finally, WISE-selected sources, called W1W2-dropouts (Eisenhardt et al. 2012) are also included. These are faint or undetected in the 3.4 and 4.6$\mu$m WISE bands but clearly detected at 12 and 22 $\mu$m --- they are also known as hot dust obscured galaxies (hot DOGs; Wu et al. 2012\nocite{Wu12}; Jones et al. 2014\nocite{Jones14}; Tsai et al. 2015\nocite{Tsai15}, Fan et al. 2016\nocite{Fan16}). 

\begin{figure*}
\epsfig{file=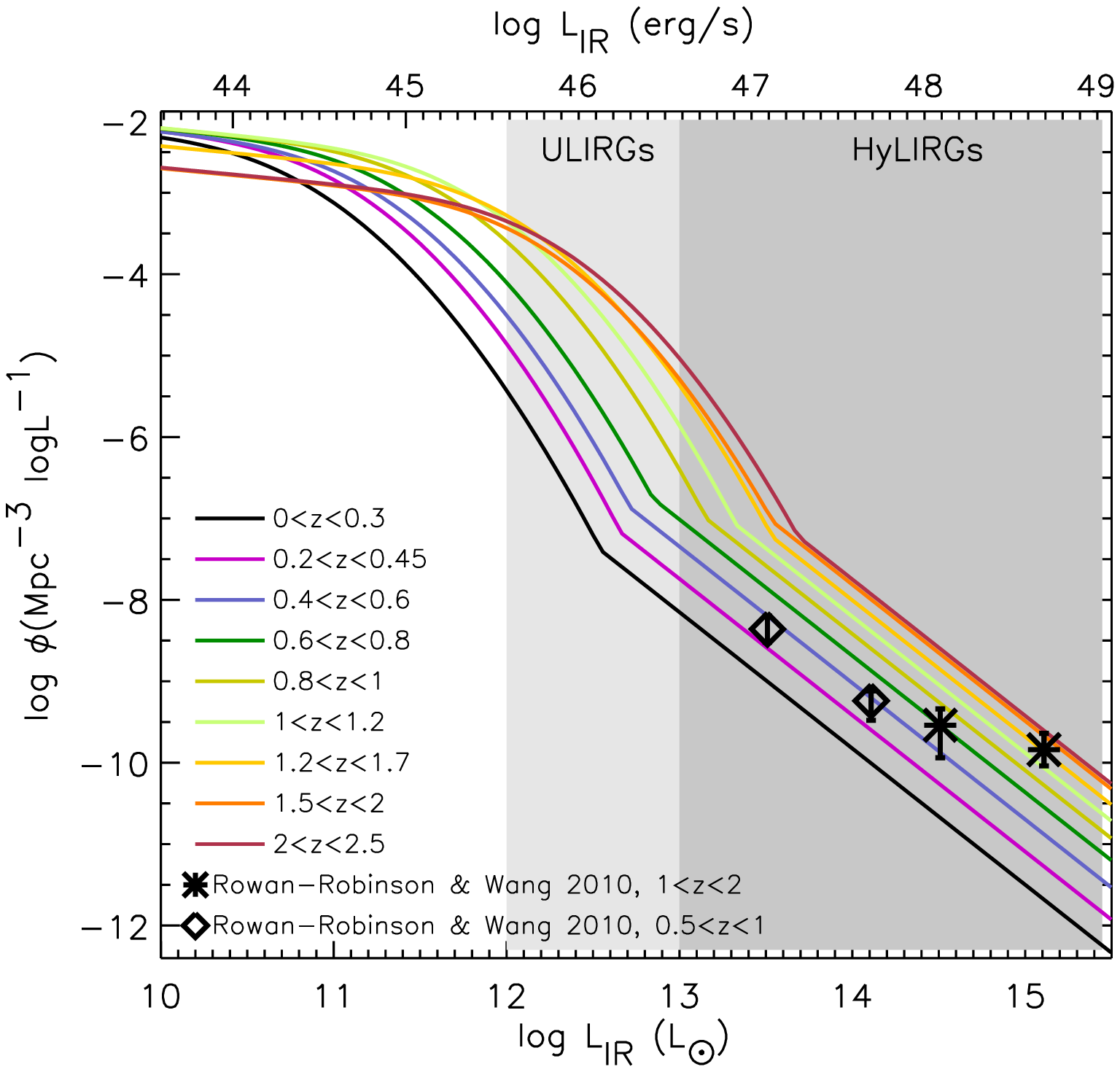,width=0.6\linewidth} 
\caption{The shape of $\phi_{\rm IR}$ over 5 orders of magnitude in $L_{\rm IR}$, for the 9 $\phi_{\rm IR}$ redshift bins shown in Fig \ref{fig:LFs}. The point showing an abrupt change in slope is $L_{\rm merge}$. At $L_{\rm IR}<L_{\rm merge}$ the LF has the shape of $\phi_{\rm IR}$ as described by the G13 models, whereas at $L_{\rm IR} \geq L_{\rm merge}$, $\phi_{\rm IR}$=$\phi_{\rm IR, AGN}$ so $\phi_{\rm IR}$ essentially assumes the slope of $\phi_{\rm IR, AGN}$ as described by the A15 models. The diamonds and asterisks show the IRAS HyLIRG luminosity function from Rowan-Robinson $\&$ Wang (2010) in the $0.5<z<1$ and $1<z<2$ ranges respectively, representing the only measurements of IR LF at those luminosities. Note that they are in agreement with our prediction of the $\phi_{\rm IR}$ slope at $L_{\rm IR} \geq L_{\rm merge}$, where the $\phi_{\rm IR}$ is essentially made up of AGN. }
\label{fig:LFtotalshape}
\end{figure*}

It is clear that the most luminous sources (currently with public data) accumulate in the transition or AGN-dominated regions, suggesting that their IR emission either has a significant AGN contribution or it is entirely dominated by the AGN, a finding which is corroborated by the studies from which they were taken. The hot DOGs are thought to be AGN powered based on several AGN signatures in the optical, mid-IR and X-rays (e.g. Wu et al. 2012; Stern et al. 2014\nocite{Stern14}; Tsai et al. 2015; Assef et al. 2015\nocite{Assef15}; Vito et al. 2018\nocite{Vito18}) and so are the QSOs. The \textit{IRAS} HyLIRGs, the same sources whose LF is shown in Fig \ref{fig:LFtotalshape} (see discussion in section \ref{sec:shape}), are unsurprisingly well within the AGN-dominated region.

\subsection{Maximum SFRs}
SFRs are thought to be proportional to $L_{\rm IR}$ (e.g. Kennicutt 1998; 2009) and hence broadband infrared photometry is often used to estimate galaxy SFRs. However, earlier we showed that the AGN contribution increases as a function of $L_{\rm IR}$, at any given redshift, suggesting that at some point $L_{\rm IR}$ will stop tracing the SFR and instead will trace the AGN power. Using the relation between $\mathcal F_{\rm AGN}$ and $L_{\rm IR}$ at each redshift bin (see Fig \ref{fig:ratio}), we compute the luminosity attributed to star-formation ($L_{\rm IR, SF}$) as follows:
\begin{equation}
L_{\rm IR, SF}=L_{\rm IR} (1-\mathcal F_{\rm AGN})
\label{eq:SFR}
\end{equation}
Subsequently, we convert $L_{\rm IR, SF}$ to SFR using the Kennicutt (1998) calibration, namely $\rm SFR=4.5 \times 10^{-44} L_{\rm IR, SF}$, where $L_{\rm IR, SF}$ is in units of erg/s. Note that since $\mathcal F_{\rm AGN}$ represents the fraction of AGN-dominated galaxies, not the fraction of AGN-powered IR emission in individual galaxies, $L_{\rm IR, SF}$ represents the amount of IR emission that can be attributed to star-formation for a galaxy population in a given redshift--$L_{\rm IR}$ bin and does not refer to individual galaxies. 

\begin{figure}
\epsfig{file=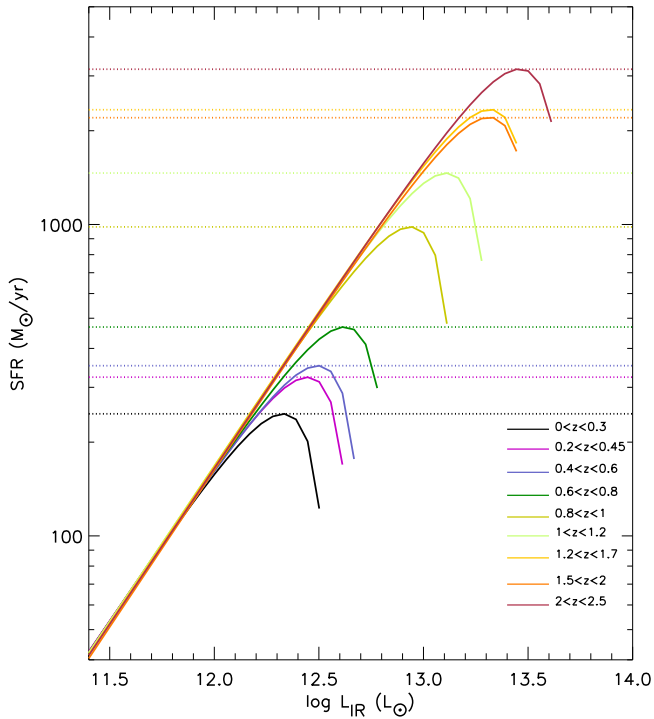,width=0.99\linewidth} 
\caption{SFR as a function of $L_{\rm IR}$ (solid lines) at each redshift bin. The SFR is computed using equation \ref{eq:SFR}. The dotted horizontal lines denote the turnover in the SFR-$L_{\rm IR}$ relation at each redshift, representing the typical maximum value of SFR that would be believable if computed from the $L_{\rm IR}$. }
\label{fig:maxSFR}
\end{figure}

Fig. \ref{fig:maxSFR} shows SFR plotted against $L_{\rm IR}$ for each redshift bin. Note that the SFR-$L_{\rm IR}$ proportionality relation breaks down at high $L_{\rm IR}$. At all redshifts, the turnover in the relation occurs approximately when $\mathcal F_{\rm AGN}\sim0.35$ (indicated by the horizontal dotted lines in Fig. \ref{fig:maxSFR}), but the $L_{\rm IR}$ at which it happens increases with increasing redshift. The turnover SFR (SFR$_{\rm turn}$), although not a hard limit, represents the typical maximum value of SFR that would be believable if computed from the $L_{\rm IR}$ at each redshift (listed in table \ref{tab:maxsfr}). Higher SFRs would likely be overestimates.  

\begin{table}
\centering
\caption{Table showing the SFR at the turnover point in the SFR-$L_{\rm IR}$ relation, i.e. the maximum believable SFR that can be computed from $L_{\rm IR}$ at each redshift bin, as shown in Fig. \ref{fig:maxSFR}. The SFRs are rounded to the nearest decade. The redshifts are quoted in the middle of the bins.}
\begin{tabular}{|l|c|c|c|c|}
\hline 
z  & SFR$_{\rm turn}$ (M$_{\odot}/yr$)\\
\hline
      0.15   &   250  \\
      0.325 &  320 \\
      0.5 &      350   \\
      0.7   &    470  \\
      0.9   &     980   \\
      1.1   &     1460    \\
       1.45   &     2340   \\
       1.75  &     2200   \\
       2.25  &      3150   \\
\hline
\end{tabular}
\label{tab:maxsfr}
\end{table}

\subsection{The shape of $\phi_{\rm IR}$ over 5 orders of magnitude in $L_{\rm IR}$}
\label{sec:shape}

As mentioned earlier, $\phi_{\rm IR} \geq \phi_{\rm IR, AGN}$ and since $\phi_{\rm IR}$ declines faster than $\phi_{\rm IR, AGN}$, there comes a point where they merge. Note that although the parametric models of the LFs in Fig \ref{fig:LFs} seem to cross-over, this is simply the effect of extrapolating them. In reality the two LFs never cross over and the condition $\phi_{\rm IR} \geq \phi_{\rm IR, AGN}$ always holds. At $L_{\rm merge}$, the space densities of AGN and galaxies become consistent within the errors, suggesting that $\phi_{\rm IR} = \phi_{\rm IR, AGN}$. Beyond $L_{\rm merge}$, $\phi_{\rm IR}$=$\phi_{\rm IR, AGN}$ still holds, hence $\phi_{\rm IR}$ should assume the slope of $\phi_{\rm IR, AGN}$ as described by the A15 models. Joining up the functional forms of $\phi_{\rm IR}$ and $\phi_{\rm IR, AGN}$ at $L_{\rm merge}$ gives the shape that the functional form of $\phi_{\rm IR}$ should have if we were able to measure it over 5 orders of magnitude in luminosity; see Fig. \ref{fig:LFtotalshape}. Note that although the change of slope at $L_{\rm IR}=L_{\rm merge}$ seems abrupt, it is because we are crudely joining the parametric forms of the two LFs at that point. If we were able to measure the space densities of sources around $L_{\rm merge}$ we would expect the change of slope to look smoother. 

The prediction that eventually $\phi_{\rm IR}$ assumes the slope of $\phi_{\rm IR, AGN}$ was first made in SP18 for $1<z<2$. Objects in the hyperluminous infrared galaxy (HyLIRG) regime ($L_{\rm IR}>10^{13}$\,L$_{\odot}$) are rare and in order to measure their space densities, an all sky survey, such as \textit{IRAS} or \textit{WISE} would be required. Using \textit{IRAS} data, Rowan-Robinson $\&$ Wang (2010; hereafter RRW10\nocite{RRW10}) estimated the HyLIRG LF at $0.5<z<1$ and $1<z<2$. SP18 showed that the HyLIRG space densities at $1<z<2$ are consistent with their predicted IR LF slope at those luminosities. Here we show that this is also the case for the HyLIRG space densities at $0.5<z<1$ (Fig. \ref{fig:LFtotalshape}). The agreement between the measured space densities of \textit{IRAS} galaxies and our modelled IR LF, confirms our prediction that at $L_{\rm IR} \geq L_{\rm merge}$, $\phi_{\rm IR}$=$\phi_{\rm IR, AGN}$ hence the IR LF is essentially made up of sources which derive the bulk of their IR power from AGN not star-formation. This suggests that the most luminous infrared emitters are AGN powered --- at least up to $z=2.5$ where our modelled $\phi_{\rm IR}$  can be compared with data, and plausibly at all redshifts.

\begin{figure}
\epsfig{file=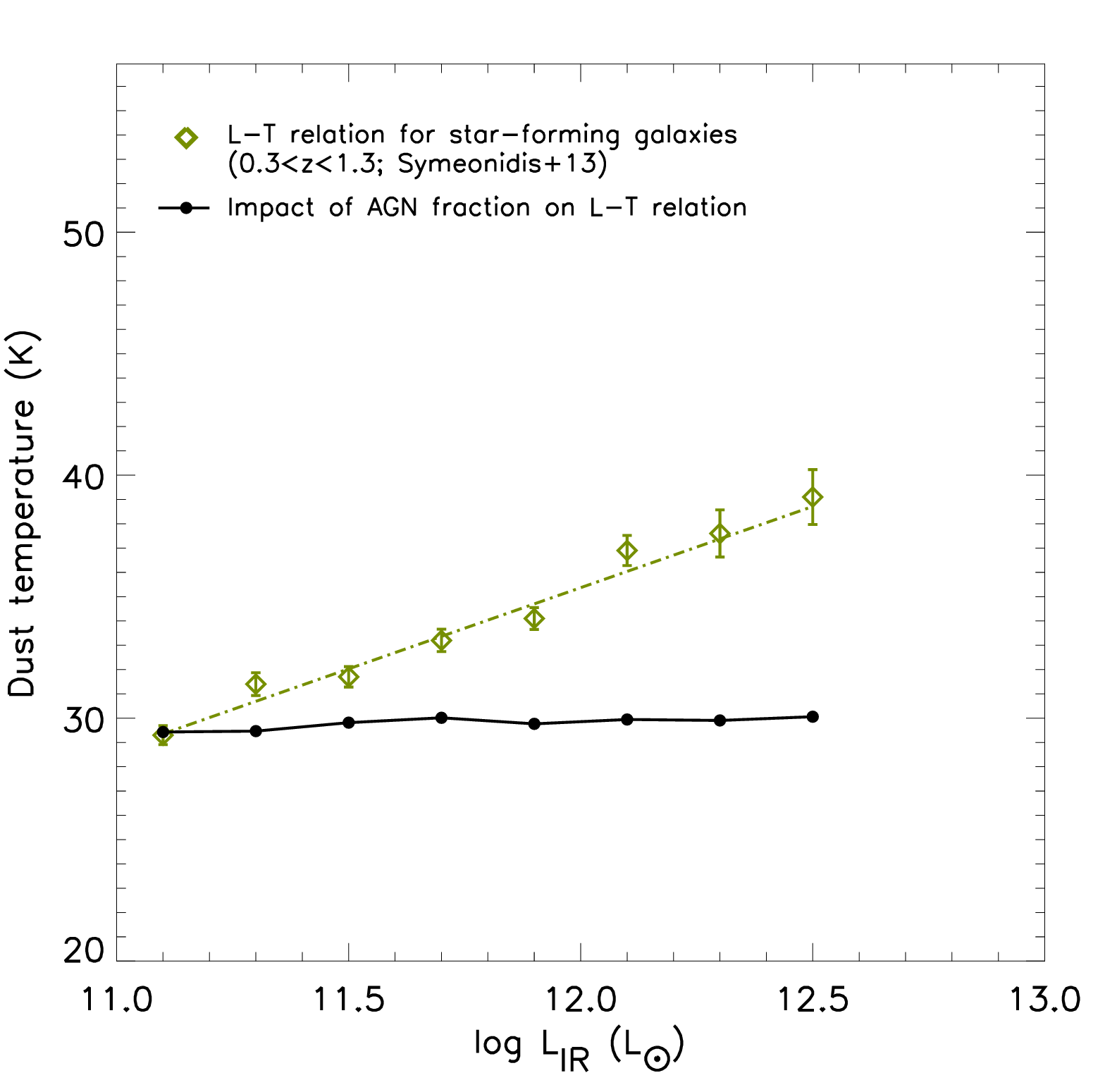,width=0.99\linewidth} 
\caption{The $L-T$ relation for $z<1.5$ star-forming galaxies taken from Symeonidis et al. (2013) (green diamonds), fitted with a straight line (dashed-dotted green line). Every bin is at a different average redshift, from left to right these are: 0.36, 0.42, 0.54, 0.67, 0.84, 0.94, 1.09, 1.23. Using equation \ref{eq:temp} and selecting $\mathcal F_{\rm AGN}$ in the appropriate redshift range, we compute the expected $T_{\rm dust}$ (black points and line), by assuming $T_{\rm dust, AGN}$=57\,K and $T_{\rm dust, SF}$=29\,K which is the dust temperature of the $z=0.36$ bin of the Symeonidis et al. (2013) relation. The flat slope of the black line reflects the fact that for all bins, $\mathcal F_{\rm AGN}$ is at its baseline level of a few percent, indicating that AGN are not responsible for the rise in temperature seen in the Symeonidis et al. (2013) $z<1.5$ $L-T$ relation, suggesting that its shape is determined by an increase in SFR from the low to the high luminosity sources. }
\label{fig:LThighz}
\end{figure}

\begin{figure}
\epsfig{file=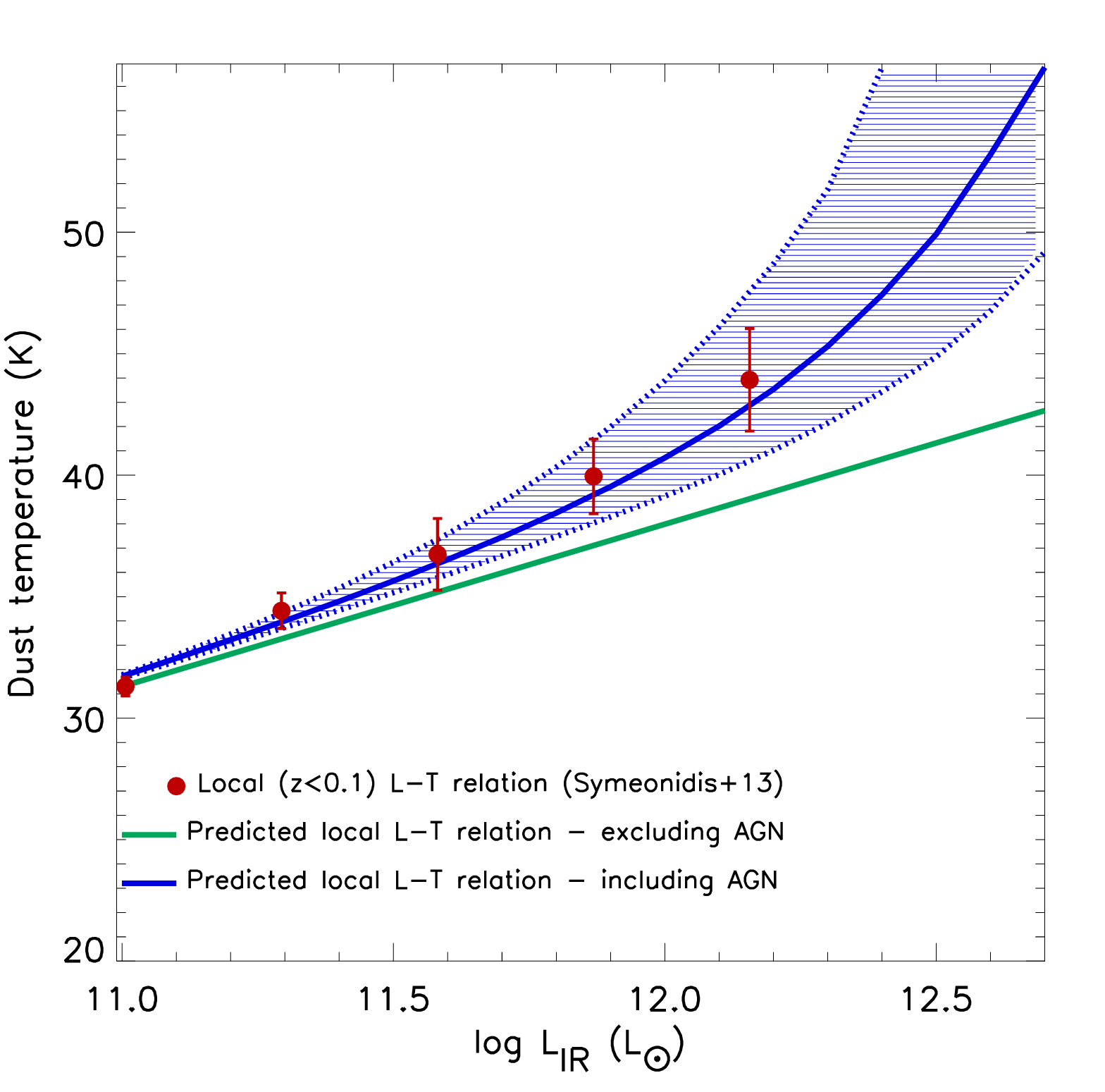,width=0.99\linewidth} 
\caption{Luminosity and dust temperature measurements for local ($z<0.1$) galaxies in five bins taken from Symeonidis et al. (2013) (red filled-in circles). The green line is the intermediate redshift ($z<1.5$) $L-T$ relation from Symeonidis et al. (2013) --- see Fig. \ref{fig:LThighz} --- normalised to the first bin of the local measurements (log\,$L_{\rm IR}/L_{\odot}$=11; 31.3\,K), and represents the local $L-T$ relation expected in the \textit{absence} of AGN, i.e. for purely star-forming galaxies. The blue curve and hatched 1$\sigma$ uncertainty represents the local $L-T$ relation expected \textit{including} AGN, computed using equation \ref{eq:temp}.  }
\label{fig:LTlocal}
\end{figure}

\subsection{The impact of AGN on the $L_{\rm IR}-T_{\rm dust}$ relation}

It is well established that there is a relation between galaxies' $L_{\rm IR}$ and their average dust temperature ($T_{\rm dust}$), with more IR-luminous systems having higher $T_{\rm dust}$  --- hereafter, we refer to this as the $L-T$ relation. This is observed in the local ($z<0.1$) Universe (e.g. Dunne et al. 2000\nocite{Dunne00}; Dale et al. 2001\nocite{Dale01}; Dale $\&$ Helou 2002\nocite{DH02}; Chapman et al. 2003\nocite{Chapman03}; Chapin et al. 2009\nocite{CHA09}) but also at higher redshifts (Hwang et al. 2010\nocite{Hwang10}; Amblard et al. 2010\nocite{Amblard10}; Calanog et al. 2013\nocite{Calanog13}; Symeonidis et al. 2013\nocite{Symeonidis13a} --- hereafter referred to as S13). The average increase of dust temperature with luminosity is often attributed to the presence of more intense starburst regions in the more luminous sources. However, SP19 showed that infrared emission in the local ULIRG population includes a substantial AGN contribution, and thus they proposed that additional dust heating by the AGN could also play a role in increasing the average dust temperatures of these systems.  

Here, we provide a simple prescription in which we use the computed $\mathcal F_{\rm AGN}$ to examine the effect of AGN dust heating as a function of $L_{\rm IR}$ in a statistical manner, and subsequently use this model to understand the local $L-T$ relation. Since our approach is based on $\mathcal F_{\rm AGN}$, as derived in section \ref{sec:AGNfrac}, it assumes a mix of AGN-dominated and star-formation dominated galaxies, rather than a scenario where AGN and star-formation emission is mixed in individual galaxies. While these two scenarios have different implications for the variations in temperature between individual galaxies at a given luminosity, we expect them to lead to similar average temperatures for the population in the wide luminosity bins we are considering. 

Since we know the AGN contribution as a function of $L_{\rm IR}$ ($\mathcal F_{\rm AGN}$; Fig \ref{fig:ratio}), the average dust temperature of galaxies can be approximated by the mixing of hot dust emission from the AGN with cooler dust emission from stellar-heated dust, using $\mathcal F_{\rm AGN}$ to weigh the AGN and star-forming galaxy dust temperatures as follows:
\begin{equation}
T_{\rm dust}=\mathcal F_{\rm AGN}T_{\rm dust, AGN}+(1-\mathcal F_{\rm AGN})T_{\rm dust, SF}
\label{eq:temp}
\end{equation}
where $T_{\rm dust, AGN}$ is the assumed dust temperature of AGN and $T_{\rm dust, SF}$ is the assumed dust temperature of star-forming galaxies. 
To obtain $T_{\rm dust}$ as a function of $L_{\rm IR}$ we assume that $T_{\rm dust, SF}$ is a function of $L_{\rm IR}$ and that $T_{\rm dust, AGN}$ is constant. To compute $T_{\rm dust, AGN}$ we measure the dust temperature of \textit{each} intrinsic AGN SED (see S17) that makes up the S16 average intrinsic AGN SED used here, by fitting a greybody function of the form $B_{\lambda}(T) \lambda^{-\beta}$ (where $\beta=1.5$) to 60 and 100$\mu$m. This wavelength range was chosen so that it is consistent with how the temperatures of local galaxies were calculated in S13; see below. Averaging these AGN SED temperatures gives $T_{\rm dust, AGN}$ of 57\,K. 

For $T_{\rm dust, SF}$ we need an $L-T$ relation for star-forming galaxies, clean from AGN contamination. For this purpose we use the $L-T$ relation in S13 derived for a sample of intermediate redshift ($z<1.5$) \textit{Herschel}-selected galaxies. Implicit in this, is the assumption that AGN do not contribute to dust heating in the S13 sample and hence the S13 $z<1.5$ $L-T$ relation is solely the result of an increase in the star-formation rate. Before using this relation, we examine whether this is indeed the case, by computing the $T_{\rm dust}$ we would expect with equation \ref{eq:temp}, assuming that $T_{\rm dust, SF}$ is constant at 29\,K which is the temperature of the first bin in the S13 $L-T$ relation. $T_{\rm dust, AGN}$ is taken to be 57\,K, as above. The results are shown in Fig. \ref{fig:LThighz}. The recomputed $z<1.5$ $L-T$ relation is flat, showing no increase with $L_{\rm IR}$ suggesting that AGN cannot be responsible for the increase in dust temperature above the assumed baseline of 29\,K. Indeed at the redshift and luminosity ranges probed by the S13 $L-T$ relation, $\mathcal F_{\rm AGN}$ is at its baseline level of a few per cent (Fig. \ref{fig:ratio}). We can thus assume that the increase in dust temperature seen in the S13 $z<1.5$ $L-T$ relation is solely a consequence of an increase in the SFR for the more luminous sources. 

As mentioned earlier, our purpose is to examine the SP19 hypothesis that AGN dust heating plays a role in shaping the local $L-T$ relation. Since we have just shown that the S13 $z<1.5$ $L-T$ relation is free from AGN contamination, we are in a position to use this as our model of what the local $L-T$ relation should look like in the absence of AGN. To do this, we first re-normalise it to the baseline temperature measured for local IR galaxies. This is 31.3\,K at log\,$L_{\rm IR}/L_{\odot}\sim10$, i.e. the first bin in the local $L-T$ relation as measured by S13 (by fitting a greybody to the 60 and 100$\mu$m data of local IR-luminous galaxies). The renormalised $L-T$ relation now represents what is \textit{expected} for the local Universe in the absence of AGN, i.e. for purely star-forming galaxies (see Fig. \ref{fig:LTlocal}). Note that the \textit{measured} dust temperatures of local sources progressively diverge from the \textit{expected} local $L-T$ relation with increasing $L_{\rm IR}$, suggesting that the increase in SFR alone cannot account for the rise in dust temperature. We now investigate whether this discrepancy is the effect of the AGN contribution to dust heating. Taking equation \ref{eq:temp} and substituting 57\,K for $T_{\rm dust, AGN}$ and the \textit{expected} local $L-T$ relation for $T_{\rm dust, SF}$, we find that $T_{\rm dust}$ is now consistent with the measured dust temperatures of local galaxies. This suggests that AGN dust heating could play a significant role in shaping the local $L-T$ relation.

Note that, as mentioned above, our model assumes that $T_{\rm dust, AGN}$ is constant, which might be an over-simplification. Indeed, $T_{\rm dust, AGN}$ may be increasing with increasing $L_{\rm IR}$, as a result of an increase in the AGN radiation power heating the dust. However to measure the empirical relationship between $T_{\rm dust, AGN}$ and $L_{\rm IR}$, a much larger AGN sample would be needed than the one available to us. In any case, an increase of $T_{\rm dust, AGN}$ with $L_{\rm IR}$ would serve to strengthen our conclusions, in the sense that it would make the role of AGN dust heating in shaping the local $L-T$ relation even more pronounced.

\begin{figure}
\epsfig{file=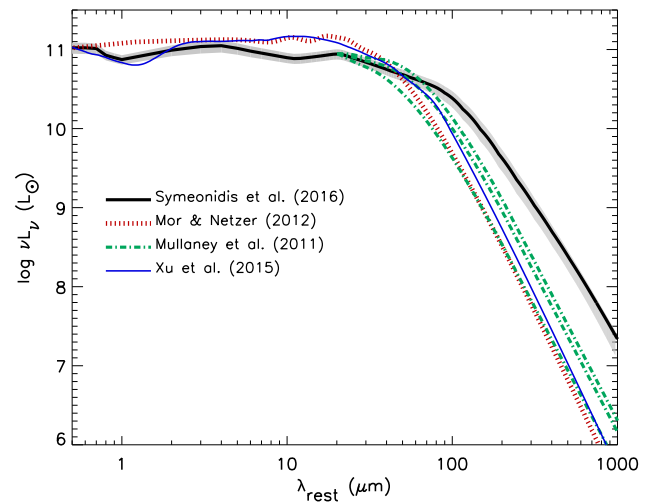,width=0.99\linewidth}\\
\caption{The Symeonidis et al. (2016) intrinsic AGN SED (black solid line) and 68 per cent confidence intervals (shaded region), compared with: (i) the AGN SEDs from Mullaney et al. (2011) normalized to the S16 SED at 20$\mu$m (dashed-dot green curves), (ii) the Xu et al. (2015) SED taken from Lyu $\&$ Rieke (2017), normalised to the S16 SED at 0.51$\mu$m (solid blue curve) and (iii) the Mor $\&$ Netzer 2012 SED extended into the far-IR as described in Netzer et al. (2016), normalised to the S16 SED at 0.51$\mu$m (vertical dash red curve). }
\label{fig:AGNseds}
\end{figure}

\begin{figure*}
\epsfig{file=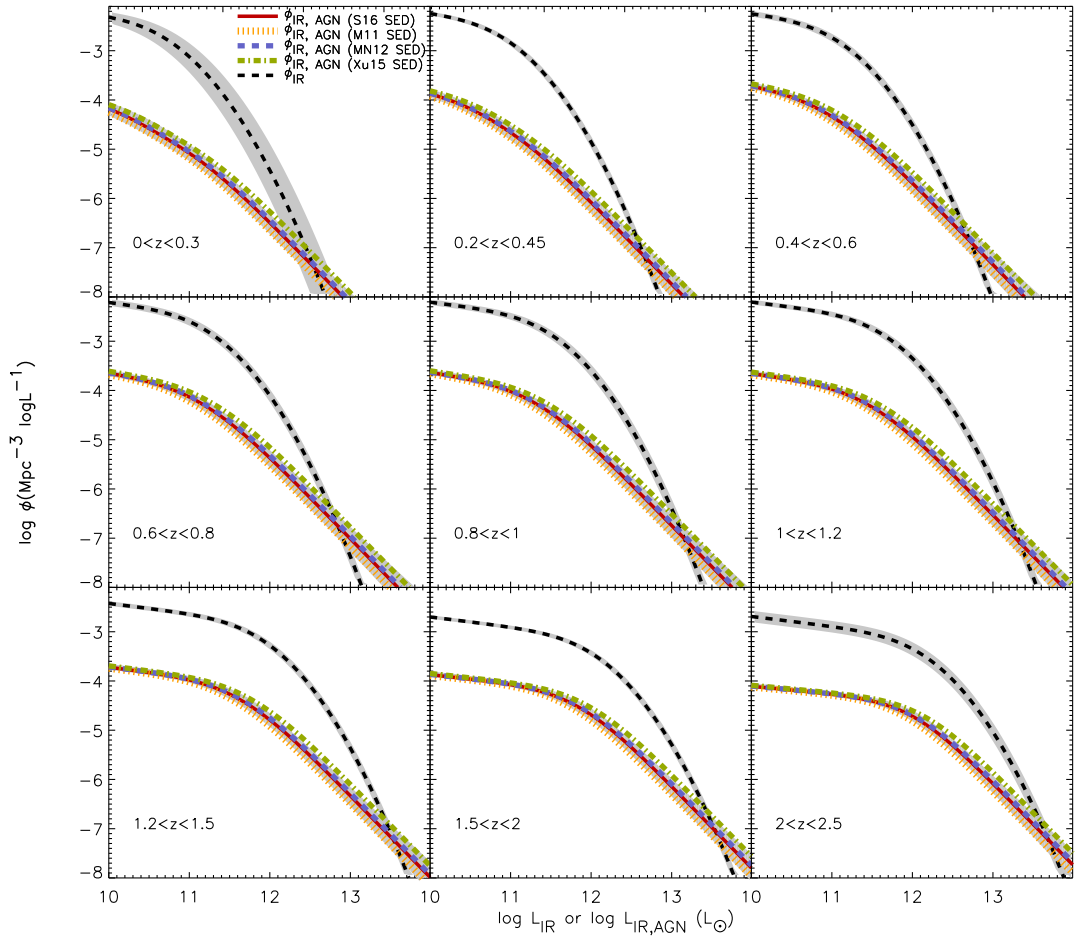,width=0.8\linewidth}\\
\caption{A modification of Fig \ref{fig:LFs} in order to show the effect of the choice of AGN SED on $\phi_{\rm IR, AGN}$. The black dashed curve and surrounding shaded region (1$\sigma$ uncertainty) is the functional form of $\phi_{\rm IR}$. The other shaded curve is the 1$\sigma$ uncertainty corresponding to $\phi_{\rm IR, AGN}$ with each coloured curve representing the functional form of $\phi_{\rm IR, AGN}$ derived with different AGN SEDs: the red solid curve corresponds to the Symeonidis et al. (2016) SED, the orange dashed curve corresponds to the Mullaney et al. (2011) SED, the blue dashed curve corresponds to the Mor $\&$ Netzer (2012) SED and the green dash-dot curve corresponds to the Xu et al. (2015) SED. The 1$\sigma$ $\phi_{\rm IR, AGN}$ uncertainty is computed with the Symeonidis et al. (2016) SED. }
\label{fig:modifiedLFs}
\end{figure*}

\begin{figure*}
\epsfig{file=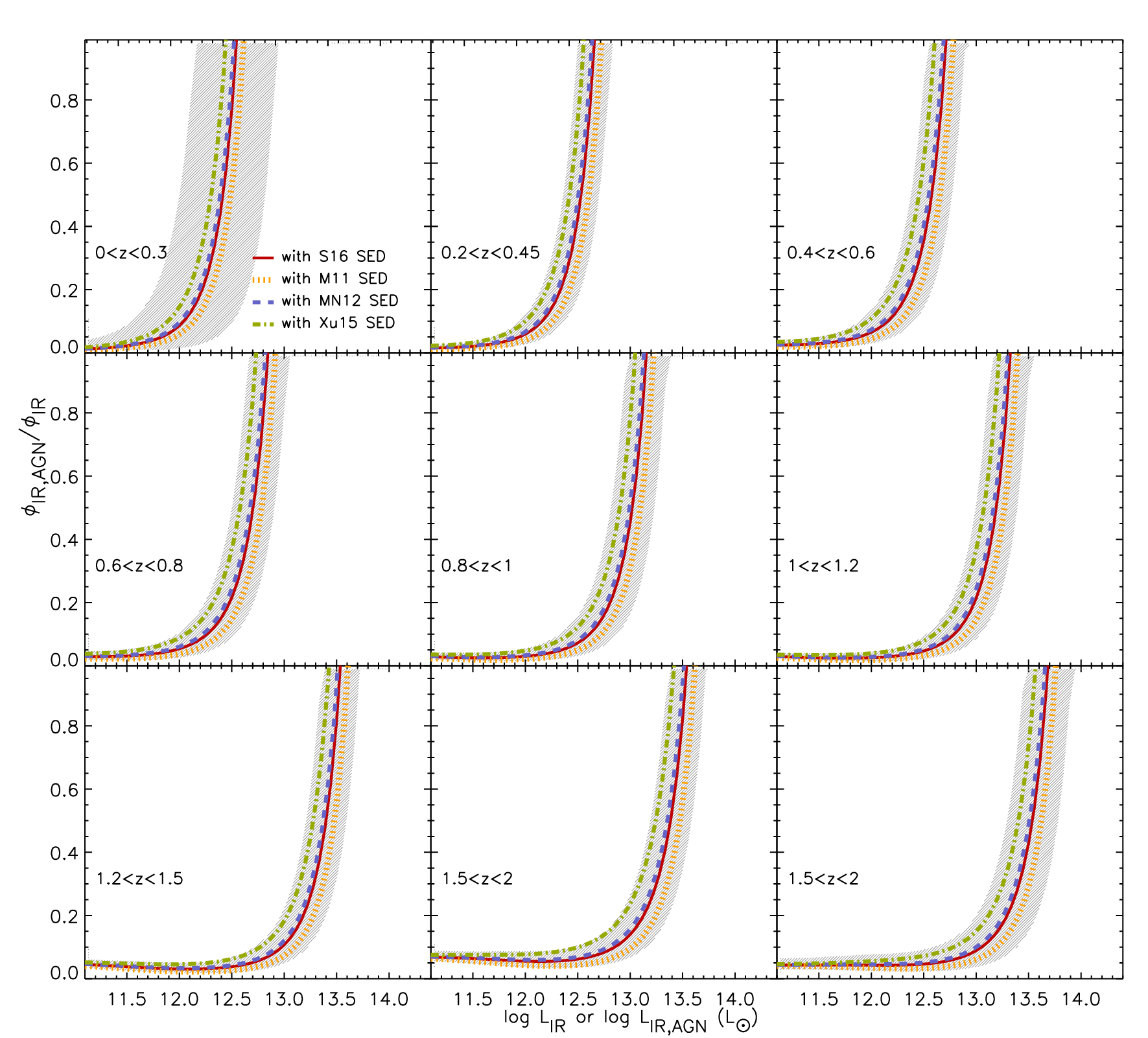,width=0.8\linewidth}\\
\caption{A modification of Fig \ref{fig:ratio} in order to show the effect of the choice of AGN SED on the ratio of $\phi_{\rm IR, AGN}$ to $\phi_{\rm IR}$ ($\mathcal F_{\rm AGN}$). The coloured curves represent $\mathcal F_{\rm AGN}$ derived using different AGN SEDs: the red solid curve corresponds to the Symeonidis et al. (2016) SED, the orange dashed curve corresponds to the Mullaney et al. (2011) SED, the blue dashed curve corresponds to the Mor $\&$ Netzer (2012) SED and the green dash-dot curve corresponds to the Xu et al. (2015) SED. The shaded 1$\sigma$ $\phi_{\rm IR, AGN}$ uncertainty is computed with the Symeonidis et al. (2016) SED. }
\label{fig:modifiedratio}
\end{figure*}

\subsection{The effect of the choice of IR AGN SED on computing $\phi_{\rm IR,AGN}$}
\label{sec:otherseds}

As mentioned in section \ref{sec:method}, the derivation of $\phi_{\rm IR, AGN}$ was based on the S16 SED. Here, we examine the impact of the choice of AGN SED, by recomputing $\phi_{\rm IR,AGN}$ with a range of diverse SEDs, chosen to be representative of the types of unobscured AGN SEDs available in the literature. These are the Xu et al. (2015) SED (hereafter Xu15 SED) taken from Lyu $\&$ Rieke (2017), the Mor $\&$ Netzer 2012 SED (hereafter MN12 SED) extended into the far-IR as described in Netzer et al. (2016), and the Mullaney et al. (2011) SEDs (hereafter M11 SEDs). All are shown in Fig. \ref{fig:AGNseds}. The MN12 and Xu15 SEDs extend from the optical to the submm and in Fig \ref{fig:AGNseds} they are shown normalised to the S16 SED at 0.51$\mu$m. One can see that, although they are less luminous in the far-IR, they are more luminous in the mid-IR and their $L_{\rm IR}$/$\nu L_{\nu, 5100}$ ratio is higher than that of the S16 SED. For the MN12 SED $L_{\rm IR}$/$\nu L_{\nu, 5100}$ =1.65 and for the Xu15 SED $L_{\rm IR}$/$\nu L_{\nu, 5100}$=2.12, compared to $L_{\rm IR}$/$\nu L_{\nu, 5100}$=1.54 for the S16 SED. On the other hand, the M11 SEDs do not extend to the optical, so in order to use them, we normalise them at 20$\mu$m to the S16 SED and assume the S16 SED shape shortwards of 20$\mu$m. This is equivalent to using the S16 SED up to 20$\mu$m with a reduced far-IR emission ($>20\mu$m), so in this way we can conveniently examine the effect of the far-IR contribution in isolation. To cover the most extreme scenario, we chose the M11 SED with the lowest far-IR emission, out of their suite of three SEDs. The $L_{\rm IR}$/$\nu L_{\nu, 5100}$ ratio for our chosen M11 SED is 1.22, indicating that reducing the far-IR emission alone only reduces $L_{\rm IR}$ by about 20 per cent. This is because more than 90 per cent of the $L_{\rm IR}$ in the aforementioned AGN SEDs (and unobscured AGN SEDs in general) is made up by emission at $\lambda<100\mu m$ (e.g. see S17). This is also the reason why the Xu15 and MN12 SEDs have higher $L_{\rm IR}$/$\nu L_{\nu, 5100}$ ratios than the S16 SED even though have lower far-IR luminosity; it is because their mid-IR luminosity is higher. 

Figs \ref{fig:modifiedLFs} and \ref{fig:modifiedratio} show the effect of the choice of AGN SED in computing $\phi_{\rm IR, AGN}$ and $\mathcal F_{\rm AGN}$ respectively. Taking the conversion with the S16 SED as the reference point, we find that other AGN SEDs introduce only a small change in our results, shifting $\phi_{\rm IR, AGN}$ and $\mathcal F_{\rm AGN}$ by about $\pm 0.1$\,dex in the abscissa and within the original $1\sigma$ uncertainties. Moreover, it is clear that the S16 AGN SED represents a middle ground within the range of available AGN SEDs. For these reasons we consider our results and conclusions robust to the choice of AGN SED.

\section{Summary and Discussion}
\label{sec:discussion}

We have compared the infrared galaxy LF ($\phi_{\rm IR}$) as a function of $L_{\rm IR}$ (i.e. bolometric 8-1000$\mu$m emission from dust heated by stars and AGN) to the infrared AGN LF ($\phi_{\rm IR, AGN}$) as a function of $L_{\rm IR, AGN}$ (i.e. bolometric 8-1000$\mu$m emission from dust heated by AGN only) up to $z=2.5$. We found that at low luminosities, $\phi_{\rm IR}$ and $\phi_{\rm IR, AGN}$ are offset by up to 2\,dex, but this difference decreases with increasing luminosity, and eventually $\phi_{\rm IR}$ and $\phi_{\rm IR, AGN}$ converge. Since the ratio of the two ($\phi_{\rm IR, AGN}$/$\phi_{\rm IR}$) is a proxy for the fraction of AGN-dominated sources ($\mathcal F_{\rm AGN}$) we found that AGN-powered galaxies constitute a progressively larger fraction of the total space density of IR-emitting sources with increasing $L_{\rm IR}$, until they take over as the dominant population. This occurs at the point when the two LFs converge, $L_{\rm merge}$. At $L_{\rm IR} \ge L_{\rm merge}$, $\phi_{\rm IR}$ assumes the slope of $\phi_{\rm IR, AGN}$: galaxies are now AGN-dominated --- true at all redshifts. However, since the LFs evolve with redshift, so does the $\mathcal F_{\rm AGN}$--$L_{\rm IR}$ relation, and at a given $L_{\rm IR}$, the fraction of AGN-dominated sources is higher at low redshift than it is at high redshift. 

Comparing the AGN and galaxy IR LFs, and their evolution with redshift has allowed us to investigate the balance of power between AGN and stars as a function of galaxy luminosity and cosmic time, and thus understand in more detail various aspects of galaxy evolution. These are discussed in more detail below.

\subsection{The shape of $\phi_{\rm IR}$}
Since the 80s, when the IR LF was first computed using \textit{IRAS} data, it has been well established that $\phi_{\rm IR}$ has a flatter high-luminosity slope than what is prescribed by the traditional Schechter (Schechter 1976\nocite{Schechter76}) shape, unlike galaxy LFs in the UV and optical. Although bright AGN are regularly removed when building UV and optical galaxy luminosity functions, this is not the case in the infrared. We thus propose that the reason for the shallower high-luminosity slope in $\phi_{\rm IR}$ at all redshifts is the increase in the space density of AGN-dominated sources relative to SF-dominated sources with increasing $L_{\rm IR}$ (first discussed in SP19 for the local Universe). We find that at $L_{\rm IR}>70L_{\star}$, the fraction of AGN-dominated galaxies rises steeply --- their relative space density increases in a given $L_{\rm IR}$ bin, flattening the slope of $\phi_{\rm IR}$. 

G13 showed that at the highest luminosities probed in their study, the IR LF is made up of galaxies which host AGN. Here we show that the IR emission in these, is in fact AGN-dominated, with the dust-reprocessed stellar emission playing a minor role. Our work is consistent with the G13 results in the sense that the most luminous galaxies \textit{must} host AGN for their emission to be AGN-dominated. In particular, the comparison between the G13 AGN LF and our $\phi_{\rm IR, AGN}$ in Fig. \ref{fig:LFsG} shows that the G13 SED fitting identifies a large proportion of luminous AGN and perhaps additional star-formation dominated sources, so the G13 AGN LF includes at least as many sources as our $\phi_{\rm IR, AGN}$.  At low AGN luminosities, particularly below the LF `knee', the comparison is perhaps not as straight-forward because AGN are often not the dominant component in the IR and so absorbed AGN might be hard to identify in the G13 SED fitting process. 

We note that the high-luminosity tail of $\phi_{\rm IR}$ undergoes a further flattening of slope when $\phi_{\rm IR}$ and $\phi_{\rm IR, AGN}$ converge at $L_{\rm IR} \sim L_{\rm merge}$. At $L_{\rm IR} \geq L_{\rm merge}$, the IR-luminous galaxies population is predominantly AGN-powered and $\phi_{\rm IR}$ is essentially shaped by the space densities of AGN. Our predictions regarding the high luminosity end are corroborated by the measured space densities of galaxies from the \textit{IRAS} all sky survey from RRW10, both at $0.5<z<1$ (this work) and $1<z<2$ (SP18); see Fig \ref{fig:LFtotalshape}. Our results indicate that if AGN did not exist, the space densities of the most luminous galaxies would be orders of magnitude lower than what is currently measured (Fig \ref{fig:LFtotalshape}), perhaps making it impossible to find such sources in our observable Universe. In other words, the reason why we are able to find galaxies at such high luminosities is because they are AGN-powered.  

Our work has thrown light on the characteristic shape of the IR LF (at $z\leq2.5$), for sources with $L_{\rm IR}>10^{10}$\,L$_{\odot}$. Moreover we have exposed that the true shape of the IR LF \textit{for star-formation} is unknown at present, although we expect it to have a steeper high luminosity slope than the IR LF currently measured, where the AGN contribution is mixed in with the star-formation contribution. Indeed cosmological simulations, in aiming to reproduce galaxy LFs, will have to take into account that the IR LF is boosted at high $L_{\rm IR}$ due to AGN. For example, Katsianis et al. (2017\nocite{Katsianis17}) find that their models require less AGN feedback to match the IR LF than to match the UV observations. This is likely because the IR LF has a shallower high-luminosity slope than the UV LF, but it is a contradictory result since AGN play a much larger role in shaping the IR LF.  

\subsection{The IR as an SFR indicator}

There are many claims in the literature that a cold dust component in galaxy SEDs is evidence for star-formation even in cases where there is a confirmed AGN, so the far-IR is often used for SFR measurements (e.g. Hatziminaoglou et al. 2010\nocite{Hatziminaoglou10}; Rosario et al. 2013\nocite{Rosario13}; Feltre et al. 2013\nocite{Feltre13}; Ellison et al. 2016\nocite{Ellison16}; Duras et al. 2017\nocite{Duras17}). The SED analysis in S16 and S17, however, indicated otherwise: it was shown that luminous enough AGN can entirely drown the IR emission of their host galaxies even in the far-IR/submm, suggesting that for sources hosting sufficiently luminous AGN, $L_{\rm IR}$ traces the AGN power rather than the star formation rate (true for all AGN types). 

Here we show that the SFR-$L_{\rm IR}$ correlation `breaks down' (Fig. \ref{fig:maxSFR}) once $\mathcal F_{\rm AGN}>0.35$. Although $\mathcal F_{\rm AGN}$ refers to the fraction of AGN-dominated sources rather than the AGN contribution in a particular galaxy, the SFR-$L_{\rm IR}$ correlation turn-over, nevertheless, implies that when a significant fraction of the population is AGN dominated in the IR, broadband IR photometry should be avoided as an indicator of star formation in individual galaxies. 

What does this imply about a limit in galaxy SFRs? By the time the AGN dominate the infrared/submm part of the electromagnetic spectrum, they are already luminous enough to dominate the bolometric emission of their host, because only about a third of the AGN power comes out in the 8-1000$\mu$m range (e.g. Tsai et al. 2015; S16; S17). As a result, the maximum believable SFRs we compute (Fig. \ref{fig:maxSFR}) might be quite close to the highest SFR that sources are likely to have at each cosmic epoch. Only with star-formation indicators independent of IR photometry will we be able to answer this conclusively. Nevertheless, our results show that `extreme starbursts' with SFRs of many thousands of M$_{\odot}$/yr are much rarer than previously thought, in agreement with cosmological models, which fail to reproduce such high SFRs even at the peak of a major merger (e.g. Narayanan et al. 2010\nocite{Narayanan10}; Narayanan et al. 2015\nocite{Narayanan15}).

\subsection{Galaxy evolution and AGN}

Galaxy evolution studies have long shown that high redshift galaxies do not as a whole have the same properties as their lower redshift counterparts for a given luminosity. This is partly because of the availability of fuel which increases with increasing redshift, but here we show that there seems to be another important factor: the AGN fraction. For a given $L_{\rm IR}$, the contribution of AGN to galaxies' energy budget decreases as a function of redshift and the typical fraction of AGN-dominated sources is higher at low redshift than it is at high redshift. This may be partly responsible for the luminosity evolution in the properties we observe in IR-luminous galaxies, i.e. high redshift ULIRGs being the analogues of low redshift LIRGs and high-redshift HyLIRGs being the analogues of low-redshift ULIRGs. 

The current picture for the formation of massive spheroidals supports a scenario whereby a major merger induces the dust enshrouded phase of intense AGN and starburst activity, followed by a blow-out phase of the dust and gas, leaving an optically unobscured QSO which eventually turns into a `dead' elliptical (e.g. Sanders et al. 1988\nocite{Sanders88a}; Hopkins et al. 2008; hereafter H08\nocite{Hopkins08}). Indeed the most IR-luminous obscured AGN, or hot DOGs as they are often referred to (see section \ref{sec:results}) are thought to be the progenitors of optically-unobscured QSOs (e.g. Assef et al. 2015; Wu et al. 2018\nocite{Wu18}) and according to Bridge et al. (2013) perhaps even the short-lived `caught-in-the-act' point where the AGN is expelling gas and dust, a claim they base on their discovery of extended Ly$\alpha$ emission in these sources. 

Our work (see Fig \ref{fig:breaklum_sources}) suggests that (i) the most luminous optically unobscured QSOs have substantial infrared emission on par with the most luminous IR-selected galaxies and (ii) the IR emission of the most luminous sources is AGN dominated, irrespective of whether the AGN is optically obscured or unobscured. This suggests that if the aforementioned scenario of evolution between optically-obscured and optically-unobscured AGN is true, it must be accompanied with a re-distribution of dust from a cocoon around the central black hole to dust extended in the AGN narrow line region (NLR). The likely existence of dust in the AGN NLR was also examined in S17 who calculated that for the AGN to retain its optical colours and produce a substantial amount of IR emission, the dust must extend over kpc scales and hence have low average dust temperature. Moreover, it was shown that the dust mass estimates for the most luminous $2<z<3$ QSOs (of the order of $10^8$M$_{\odot}$: see S17; also Ma $\&$ Yan 2015) are comparable with the dust masses calculated for hot DOGs (e.g. Fan et al. 2016). 

Since we find there are no galaxies whose luminosity from star-formation supercedes the most luminous AGN, our results are consistent with, although do not prove, the evolutionary scenario where the AGN quenches star-formation in their host galaxies. The question that remains unanswered, however, is what are the SFRs of galaxies hosting the most luminous AGN. Measuring these would be a critical step forward, if we are to understand AGN feedback in the context of galaxy evolution. Indeed, JWST will give the opportunity to explore other indicators of star-formation such as polycyclic aromatic hydrocarbons (PAHs; F{\"o}rster Schreiber et al. 2004\nocite{Forster-Schreiber04}; Peeters et al. 2004\nocite{Peeters04}; Risaliti et al. 2006\nocite{Risaliti06}; Kennicutt et al. 2009\nocite{Kennicutt09}), in the effort to move away from SFR measurements using broadband infrared photometry.

\subsection{The impact of AGN on the $L_{\rm IR}-T_{\rm dust}$ relation}

Many studies have reported the existence of intermediate redshift galaxies with lower $T_{\rm dust}$ than local galaxies of equivalent luminosities (e.g. Chapman et al. 2005\nocite{Chapman05}; Coppin et al. 2008\nocite{Coppin08}; Symeonidis et al. 2009\nocite{Symeonidis09}; Hwang et al. 2010\nocite{Hwang10}; Magnelli et al. 2012\nocite{Magnelli12}; Casey et al. 2012\nocite{Casey12}; S13). Although in some cases their detection rate is linked to the selection biases of submm surveys (e.g. Symeonidis et al. 2011\nocite{Symeonidis11a}) and hence does not reflect the average properties of the population, there is a measured $\sim$10\,K difference between the average dust temperatures of local ULIRGs and their intermediate redshift counterparts (see S13). In other words, the intermediate redshift $L-T$ relation is flatter than the local one. It has been suggested before that this is a consequence of the local $L-T$ relation evolving with redshift (e.g. Chapman et al. 2002; Lewis et al. 2005; Chapin et al. 2009) in the same way that $L_{\star}$ evolves. Symeonidis et al. (2009) looked into this claim by de-evolving the luminosities of high-z ULIRGs by $(1+z)^3$, finding that they remained outside the local $L-T$ relation. They concluded that this style of evolution of the $L-T$ relation cannot be responsible for the increased presence of cold ULIRGs in the distant Universe. 

Recently, SP19 proposed the idea that the local $L-T$ relation could be driven by AGN. Here we demonstrate that this is indeed the case. Folding in the AGN contribution to the $L-T$ relation for star-forming galaxies, by mixing the AGN dust temperature and star-forming galaxy dust temperature in the ratio prescribed by $\mathcal F_{\rm AGN}$, we successfully reproduce the $L-T$ relation measured in the local Universe (Fig \ref{fig:LTlocal}). We find that the dust temperatures of local ULIRGs are higher than would be expected if the increase in SFR were the sole factor, hence we conclude that the difference in average dust temperature between local ULIRGs and their less luminous counterparts is partly due to an increased AGN fraction in the former group. Since $\mathcal F_{\rm AGN}$ increases as a function of $L_{\rm IR}$ in a similar fashion (at least up to $z\sim2.5$), we propose that the reason why the $L-T$ relation at high redshift is flatter than the local $L-T$ relation is because of a change in the fraction of AGN-dominated galaxies. In other words, high redshift ULIRGs are cooler than their local counterparts because they are predominantly star formation dominated, in contrast to local ULIRGs, many of which are AGN-powered (see Fig. \ref{fig:breaklum_sources}). Note that this result does not contradict the observed changes in galaxy properties, such as sizes (e.g. Tacconi et al. 2006; Iono et al. 2009; Rujopakarn et al. 2011) as these are linked to the availability of cold gas which is more substantial at high redshift.

\section{Conclusions}
\label{sec:conclusions}

We have described a phenomenological approach aimed towards understanding the balance of power between AGN and stars up to $z\sim4$. Using the X-ray (converted to IR) AGN LF and the IR galaxy LF, we have investigated the impact of AGN in measured galaxy properties such as $L_{\rm IR}$, $SFR$ and $T_{\rm dust}$ and interpreted the shape of two key observables: the IR LF and the $L-T$ relation. 

Our major findings and conclusions are listed below:
\begin{itemize}
\item For the first time, we derive the shape of the pure AGN LF in the IR finding that it is responsible for shaping the IR LF at the highest luminosities. We find that star-forming galaxies dominate the IR LF at $L_{\rm IR} < 70 L_{\star}$, whereas AGN start flattening the high-luminosity tail at $L_{\rm IR} \gtrsim70 L_{\star}$. There is a further break in slope at $L_{\rm IR} \gtrsim120 L_{\star}$where the IR LF becomes AGN-dominated at which point it takes on the slope of the AGN LF. 
\item The most IR-luminous and thus bolometrically luminous sources at all redshifts are AGN-dominated. This, moreover, explains the reason why we can observe such luminous galaxies. If AGN did not exist, the space densities of $L_{\rm IR}>10^{13}$\,L$_{\odot}$ galaxies would be orders of magnitude lower than what is currently measured, rendering such sources very difficult to find in our observable Universe. 
\item The range of maximum SFRs is likely between 1000 and 4000 M$_{\odot}$/yr at the peak of cosmic star formation history ($1<z<3$). Galaxies with reported uncorrected SFRs significantly offset from this, will have a significant contribution from AGN, so the SFRs will likely need to be corrected. This casts doubt on the abundance of `extreme' starbursts. Objects claimed to be such, are most likely AGN-powered.  
\item The AGN contribution in the IR can account for the differences in average dust temperatures between sources of comparable luminosity at different redshifts. Local ULIRGs are hotter than their intermediate redshift counterparts because the AGN contribution in the former is up to 30 times larger.  
\end{itemize}

\clearpage
\section*{Acknowledgments}
MS and MJP acknowledge support by the Science and Technology Facilities Council [ST/S000216/1].

\section*{Data Availability}
The data underlying this article are either available in cited works in the article, in the article itself or will be shared on reasonable request to the corresponding author.

\bibliographystyle{mn2e}
\bibliography{references}

\end{document}